\documentclass[pra,twocolumn,aps,floatfix,showpacs,tightenlines,superscriptaddress,amsmath,amssymb,showkeys,10pt,nofootinbib,longbibliography]{revtex4-1}
\usepackage[T1]{fontenc}
\usepackage[latin1]{inputenc}
\usepackage{amsmath,graphicx}
\usepackage{amssymb}
\usepackage{epsfig,graphicx}
\usepackage{mathrsfs}
\usepackage[active]{srcltx}
\usepackage{mdframed}
\usepackage{mathtools}
\usepackage{xcolor}
\usepackage{hyperref}
\usepackage{braket}
\usepackage{gensymb}
\usepackage{enumerate}
\usepackage{epsfig,libertine,dsfont,amsthm}
\usepackage[libertine]{newtxmath}
\usepackage{layouts}
\usepackage[justification=justified]{subcaption}
\usepackage{tikz}
\usetikzlibrary{quantikz}
\usepackage[newcommands]{ragged2e}
\usepackage{hyperref}
\hypersetup{colorlinks=true,urlcolor=red,citecolor=green,filecolor=magenta}

\newcommand{\C}{\mathbb{C}}    

\newcommand{\spc}[1]{\mathcal{#1}}    

\newcommand{\Lin}{\mathsf{Lin}}       
\def\>{\rangle}                                       
\def\<{\langle}                                        
\def\kk{\>\!\>}                                          
\def\bb{\<\!\<}                                          

\newcommand{\map}[1]{\mathscr{#1}}            
\newcommand{\Tr}{\operatorname{Tr}}          

\newtheorem{theorem}{Theorem}

\newtheorem{lem}{Lemma}
\newtheorem{pror}{Proposition}
\newtheorem{cor}{Corollary}
\newtheorem{definition}{Definition}

\usepackage{tabularx}

\newcounter{protocol}
\makeatletter
\newenvironment{protocol}[1]{
  \refstepcounter{protocol}%
    \protected@edef\@currentlabelname{#1}
    \par\addvspace{\topsep}
   \noindent
   \tabularx{\linewidth}{@{} X @{}}
    \hline
    \textbf{Protocol \theprotocol} #1 \\
    \hline
    }
  { \\
   \endtabularx
   \par\addvspace{\topsep}}
\makeatother

\def\Proof{{\bf Proof.~}}
\def\qed{$\blacksquare$ \newline}

\begin{document}

\title{Dimension-independent weak value estimation via controlled {\tt SWAP} operations} 

\author{Giulio Chiribella}
\email{giulio@cs.hku.hk}
\affiliation{QICI Quantum Information and Computation Initiative, Department of Computer Science, The University of Hong Kong, Pokfulam Road, Hong Kong}
\affiliation{Quantum Group, Department of Computer Science, University of Oxford, Wolfson Building, Parks Road, Oxford, OX1 3QD, United Kingdom}
\affiliation{Perimeter Institute for Theoretical Physics, 31 Caroline Street North, Waterloo, Ontario, Canada}

\author{Kyrylo Simonov}
\email{kyrylo.simonov@univie.ac.at}
\affiliation{Fakult\"at f\"ur Mathematik, Universit\"at Wien, Oskar-Morgenstern-Platz 1, 1090 Vienna, Austria}

\author{Xuanqiang Zhao}
\email{xqzhao7@connect.hku.hk}
\affiliation{QICI Quantum Information and Computation Initiative, Department of Computer Science, The University of Hong Kong, Pokfulam Road, Hong Kong}

\date{\today}

\begin{abstract}
Weak values of quantum observables are a powerful tool for investigating  quantum phenomena. Some methods for measuring weak values in the laboratory require weak interactions and postselection, while others  are deterministic, but require statistics over a number of experiments that   grows linearly with the dimension of the measured system in the worst case over all possible observables. Here we propose a deterministic dimension-independent scheme for estimating weak values of arbitrary observables. The scheme is based on  controlled {\tt SWAP} operations, and associates  states and  observables in the mathematical expression of the weak value  to   preparations devices and measurements devices in the experimental setup, respectively. Thanks to this feature, it provides insights into the relation between    states of two identical quantum systems at a single moment of time and states of a single quantum system at two moments of time, also known as two-time states. Specifically, our scheme provides an alternative expression for  two-time states, and establishes a link between two-time states accessible through the controlled-{\tt SWAP} scheme and bipartite quantum states with positive partial transpose. 
\end{abstract} 

\maketitle 

\section{Introduction}

In a seminal 1988 paper \cite{Aharonov1988}, Aharonov, Albert, and Vaidman introduced the notion of weak values  and  showed  that they can be experimentally accessed by letting  the measured system interact weakly with a pointer in the time interval between a pre-selection and a post-selection.  Since then,   weak values have proven a powerful tool for analyzing a broad spectrum of quantum phenomena \cite{aharonov2010time,shikano2012theory, kofman2012nonperturbative,Dressel2014}.  On the foundational side, they provide a lens for  experimentally  investigating quantum paradoxes  \cite{Aharonov2002, lundeen2009experimental, Kocsis2011} and the behaviour of quantum systems in time \cite{Williams2008, dressel2011experimental,Goggin2011,groen2013partial}, and they also serve as  a quantitative indicator of   non-classicality~\cite{Pusey2014, Dressel2014, Dressel2015, Kunjwal2019}.  On the applied side,  they  provide probabilistic amplification techniques for quantum metrology  \cite{Hosten2008, Dixon2009, Brunner2010, Starling2010precision, starling2010continuous, hofmann2011uncertainty, hofmann2012estimation, strubi2013measuring} and  for the direct measurement of quantum states \cite{Lundeen2011, lundeen2012procedure, salvail2013full, malik2014direct}.

Several  experimental schemes for measuring weak values have been proposed and demonstrated in the laboratory. The first schemes, based on the original definition of weak values, involve weak measurement interactions and post-selection (see \cite{Dressel2014} for a review).   More recently, there has been a growing interest in methods for estimating weak values without post-selection and weak couplings  \cite{Johansen2007, Hiley2012, Vallone2016, Zhang2016, Denkmayr2017, Cohen2018, Denkmayr2018, Lostaglio2022, Wagner2023}. These schemes are conceptually interesting because they disentangle the notion of weak value from the notion of weak measurements,  and expand the scope of application of the notion of weak value by linking it to fundamental protocols in quantum information and computation.  For example, an ingenious method was developed by Hoffman \cite{Hoffman2012}, who showed that weak values can be estimated by performing standard measurements on the outputs of the universal quantum cloning machine \cite{Buzek1996, Gisin1997, Werner1998}.   Another  method was proposed by Wagner {\em et al}  \cite{Wagner2023},  based on a quantum algorithm called  the  cyclic shift test \cite{Oszmaniec2021}. 

A limitation of the existing deterministic estimation protocols, however,  is that the number of samples  needed to accurately estimate the weak value generally grows   linearly in the system dimension $d$.  For example, the cloning method \cite{Hoffman2012}  provides an estimate of the weak value multiplied by a term of order $1/d$.   Hence, obtaining a reliable estimate requires a number of repetitions of the experiment growing  as $d$. For a quantum system composed of $n$ particles, this scaling results into an exponential increase of the sample complexity, that is, the  number of samples needed to achieve a desired level of accuracy of estimation. A similar issue arises in the cyclic test approach     \cite{Wagner2023}, whose  sample complexity  grows with $d$ in the worst case over all possible observables.   

In this paper, we provide a deterministic  method for estimating weak values with dimension-independent sample complexity.   In our method, two identical quantum systems are prepared in the initial and final states  appearing in the definition of weak value.    Then, the two systems undergo a controlled-{\tt SWAP} operation, which exchanges them or leaves them unchanged depending on the quantum state of a control system, as illustrated in Figure \ref{fig:circuits} .  After the controlled-{\tt SWAP} operation, the two systems undergo a possibly noisy measurement, whose  outcomes are used to estimate the weak value.   

The controlled-{\tt SWAP} scheme studied in this paper  also provides insights into the theory of two-time states, a generalized type  of states that describe quantum systems subject to both pre- and post-selections \cite{Aharonov1964, Aharonov1991, Silva2014}.     We show that the expectation values associated to two-time states can be obtained from the  expectation values associated to bipartite density matrices, by applying a linear fractional transformation that involves  a {\tt SWAP} operation and a partial transpose.     Using this result,  we show that the  controlled-{\tt SWAP} method can be used to estimate the expectation values of all two-time states associated to density matrices with positive partial transpose   (PPT) \cite{Peres1996, Horodecki1996}.    Based on this fact, we discuss several adaptations of the controlled-{\tt SWAP} method that allow to access the expectation values for  two-time states associated to non-PPT states.

The rest of the paper is structured as follows. In Section  \ref{sec:EstWV}, we briefly review the notion of weak value and put forward a dimension-independent estimation scheme based on controlled-{\tt SWAP} operations. In Section \ref{sec:DoubleWV}, we develop several generalizations of our  scheme and introduce a new quantity dubbed  the ``double weak value''. In Section \ref{sec:TwoTimeStates}, we provide the reader with an overview of the theory of two-time states. The matrix representation of two-time states is discussed in Section \ref{sec:MatrRep}. In Section \ref{sec:TwoTimeSWAP}, we highlight a fundamental connection between two-time states and the developed estimation scheme. Finally, we provide conclusions  and an outlook  in Section \ref{sec:concl}. 
 
\section{Weak value estimation}\label{sec:EstWV}

In this section we provide a brief introduction to weak values, and present a deterministic scheme for estimating weak values with dimension-independent sample complexity.  

\subsection{Introduction to weak values}

Weak values were introduced by Aharonov, Albert and Vaidman, as part of a framework describing pre- and post-selected ensembles \cite{Aharonov1988}. Their original definition referred to the scenario where a quantum system $S$ is pre-selected and post-selected in two pure states, described by rays in the system's Hilbert space $\spc H$. Given two (generally unnormalized) vectors $|\psi_{\rm in}\>\in \spc H$ and $|\psi_{\rm fin}\>\in  \spc H$ satisfying the condition $\<\psi_{\rm in}|\psi_{\rm fin}\>\not =  0$, the weak value of an observable  $A$ is  defined as
\begin{align}\label{pureWV}
W (A \, |\,   \psi_{\rm in},  \psi_{\rm fin} ) : =  \frac{ \<\psi_{\rm fin}|  A  |\psi_{\rm in}\> }{\<\psi_{\rm fin}|  \psi_{\rm in}\>} \, .
\end{align}
Here the vector $|\psi_{\rm in}\>$ ($|\psi_{\rm fin}\>$) represents the initial (final) state of the system. The observable $A$ is typically taken to be a self-adjoint operator,  but more generally could be any linear operator acting on $\spc H$.  Hereafter, the algebra of all linear operators on $\spc H$ will be  denoted by $\Lin (\spc H)$.   

More recently, the notion of weak value was  generalized to mixed states  \cite{Silva2014,Vaidman2017}: for a pair of density matrices $\rho_{\rm in}$ and $\rho_{\rm fin}$ satisfying the condition $\Tr[ \rho_{\rm in}  \rho_{\rm fin}]\not =  0$, the weak value   is defined as 
\begin{align}\label{mixedWV}
W    (  A \,|\,  \rho_{\rm in},  \rho_{\rm fin} )  =  \frac{   \Tr [ \rho_{\rm fin}\, A\,  \rho_{\rm in} ]  }{\Tr[\rho_{\rm fin} \,  \rho_{\rm in}]} \, .
\end{align}
Eq. (\ref{pureWV})  can be obtained as a special case of Eq. (\ref{mixedWV}) by setting $\rho_{\rm in} =  |\psi_{\rm in}\>\<\psi_{\rm in}|$ and $\rho_{\rm fin} =  |\psi_{\rm fin}\>\<\psi_{\rm fin}|$.   Physically, the mixed-state weak value (\ref{mixedWV}) can be reduced to the pure-state weak value (\ref{pureWV}) through a purification procedure  presented in Ref. \cite{Vaidman2017}.

\begin{figure*}
\centering
\begin{subfigure}{0.22\textwidth}

\begin{quantikz}
\tikzset{
my label/.append style={above right,xshift=-0.1cm,yshift=0.1cm},
rightinternal label/.append style={xshift=1cm,yshift=1.5cm}
}\lstick{$\color{blue}{\rho_{\rm in}}$} & \swap{1} & \meter[style={draw=red}]{$\color{red}{A}$}
 \\
\lstick{$\color{blue}{\rho_{\rm fin}}$} & \targX{} & \qw \\
\lstick{$|+\rangle\langle +|$} & \ctrl{-2} & \meter{$X/Y$}
\end{quantikz}
   \caption{}
    \label{fig:first}
\end{subfigure}
\hfill
\begin{subfigure}{0.22\textwidth}
    \begin{quantikz}
\tikzset{
my label/.append style={above right,xshift=-0.1cm,yshift=0.1cm},
rightinternal label/.append style={xshift=1cm,yshift=1.5cm}
}\lstick{$\color{blue}{\rho_{\rm in}}$} & \swap{1} & \meter[style={draw=red}]{$\color{red}{A}$}
 \\
\lstick{$\color{blue}{\rho_{\rm fin}}$} & \targX{} & \meter[style={draw=red}]{$\color{red}{B}$} \\
\lstick{$|+\rangle\langle +|$} & \ctrl{-2} & \meter{$X/Y$}
\end{quantikz}
    \caption{}
    \label{fig:second}
\end{subfigure}
\hfill
\begin{subfigure}{0.22\textwidth}
    \begin{quantikz}
\tikzset{
my label/.append style={above right,xshift=-0.1cm,yshift=0.1cm},
rightinternal label/.append style={xshift=1cm,yshift=1.5cm}
}
\lstick[wires=2]{$\color{blue}{\rho_{\rm in, fin}}$} & \swap{1} & \meter[style={draw=red}]{$\color{red}{A}$}
 \\
 & \targX{} & \meter[style={draw=red}]{$\color{red}{B}$} \\
\lstick{$|+\rangle\langle +|$} & \ctrl{-2} & \meter{$X/Y$}
\end{quantikz}
    \caption{}
    \label{fig:third}
\end{subfigure}
\hfill
\begin{subfigure}{0.245\textwidth}
    \begin{quantikz}
\tikzset{
my label/.append style={above right,xshift=-0.1cm,yshift=0.1cm},
rightinternal label/.append style={xshift=1cm,yshift=1.5cm}
}
\lstick[wires=3]{$\color{blue}{\rho_{\rm in, fin}}$} & \swap{1} & \qw[style={xshift=1.09cm, yshift=0.25cm, color=red}]{{\displaystyle O}} \rstick[wires=3]{}
 \\ [-15pt] &&&|[meter, style={draw=red}]| \\ [-15pt]
 & \targX{} & \qw \\
\lstick{$|+\rangle\langle +|$} & \ctrl{-2} & \qw & \meter{$X/Y$}
\end{quantikz}
    \caption{}
    \label{fig:fourth}
\end{subfigure}
        
\caption{Controlled-{\tt SWAP} protocol for the estimation of weak values and some of its generalizations for the estimation of expectation values associated to two-time states.    (\ref{fig:first})   Estimation of the weak value $W(A|  \rho_{\rm in},  \rho_{\rm fin} ) := {\Tr[  \rho_{\rm fin}  \,A\, \rho_{\rm in} ]}/{\Tr[\rho_{\rm fin} \, \rho_{\rm in}]}$, with initial state $\rho_{\rm in}$, final state  $\rho_{\rm fin}$, and  observable $A$.    Two copies of the system are initially prepared in the states  $\rho_{\rm in}$ and   $\rho_{\rm fin}$,  undergo a controlled-{\tt SWAP} gate, with the control qubit in the  state $|+\>  = {(|0\>  +  |1\>)}/{\sqrt 2}$. Finally, a (possibly noisy) measurement of the observable $A$ is performed on the first copy, while the control is randomly measured in one of the two bases $\{|+\> ,  |-\>\}$ and $\{|+i\> ,  |-i\>\}$, consisting of the eigenstates of the Pauli observables $X$ and $Y$, respectively.  (\ref{fig:second})    Estimation of the ``double weak value'' $W_2(A, B|  \rho_{\rm in},  \rho_{\rm fin} ) := {\Tr[ A\, \rho_{\rm in}\,B\,  \rho_{\rm fin} ]}/{\Tr[\rho_{\rm fin} \, \rho_{\rm in}]}$, with observables $A$ and $B$.     (\ref{fig:third})   Variant of the protocol involving a general bipartite state $\rho_{\rm in, fin}$.  When the density matrix $\rho_{\rm in, fin}$ has positive partial transpose, this variant can be used to estimate the expectation values associated to a class of two-time states.    (\ref{fig:fourth})   Variant of the protocol involving a general bipartite state $\rho_{\rm in, fin}$ and a general bipartite observable $O$. }
\label{fig:circuits}
\end{figure*}
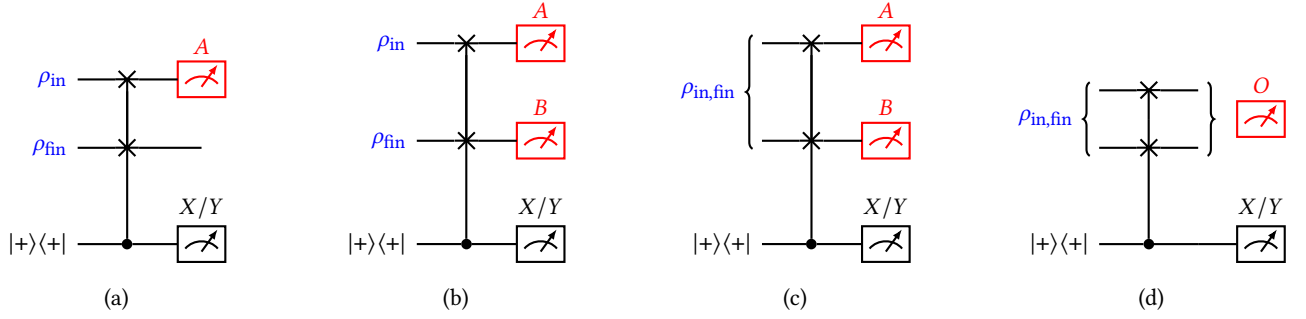

An important question is how to measure  weak values. In their seminal paper \cite{Aharonov1988},  Aharonov, Albert, and Vaidman provided an experimental scheme using weak interactions and postselection.  A number of other schemes was subsequently devised by other authors \cite{Ritchie1991, Pryde2005, Hosten2008, Dixon2009, Brunner2010, Goggin2011, Lundeen2011, Li2011, Dressel2012, Xu2013, MaganaLoaiza2014, Piacentini2016, Vaidman2017Review, Cohen2018, Rebufello2021, Adam2023, Wiseman2023}. 

The most recent scheme was presented in Ref. \cite{Wagner2023}. In its simplest version, the scheme  is defined for pure states $\rho_{\rm in} =  |\psi_{\rm in}\>\<\psi_{\rm in}|$, $\rho_{\rm fin}=|\psi_{\rm fin}\>\<\psi_{\rm fin}|$, and rank-one observables $A  =  |\alpha\>\<\alpha|$. Here, all the vectors $|\psi_{\rm in}\>, |\psi_{\rm fin} \>,$ and $|\alpha\>$ are taken to have unit length.      The protocol  consists of two subprotocols:  a cyclic test \cite{Oszmaniec2021} for estimating the trace $\Tr[ \rho_{\rm fin}\,  |\alpha\>\<\alpha|\,   \rho_{\rm in} ]$, and a {\tt SWAP} test \cite{Barenco1997, Buhrman2001} for estimating the trace $\Tr[\rho_{\rm fin}   \rho_{\rm in}]$.   The extension to mixed states is straightforward: since both the cyclic and {\tt SWAP} tests are linear in the density matrix, they can be used to estimate the traces  $\Tr[ \rho_{\rm fin} \,  |\alpha\>\<\alpha|\,    \rho_{\rm in} ]$  and  $\Tr[\rho_{\rm fin} \,    \rho_{\rm in}]$ for arbitrary $\rho_{\rm in}$ and $\rho_{\rm fin}$.    The extension to  general self-adjoint  observables $A$ is less direct, because, in the cyclic test, the rank-one observable $ |\alpha\>\<\alpha|$ is treated as a quantum state, and, therefore, it cannot be directly replaced by a general  observable. Instead, one can use any diagonalization of the operator   $A$, of the form $A =  \sum_{i=1}^r\,  a_i\,  |\alpha_i\>\<\alpha_i|$ where $r$ is the rank of $A$,  $(a_i)_{i=1}^r$ are (possibily degenerate) non-zero eigenvalues, and $\{  |\alpha_i\>\}_{i=1}^r$ is an orthonormal basis for the support of $A$.  Using this decomposition, one can then compute the weak value of $A$ as a linear combination  of the weak values of the rank-one observables  $A_i  :=  |\alpha_i\>\<\alpha_i|$.    A limitation of this approach, however, is that in general the rank of the  observable $A$ can be as large as $d$,  and therefore the number of experimental settings needed to estimate the weak value of $A$ grows with the system's dimension in the worst case.  For a system of $n$ particles, the number of settings for the estimation of a generic weak value grows exponentially with $n$.

\subsection{The controlled-{\tt SWAP} protocol}

We now provide a way to estimate weak values of arbitrary observables in a dimension-independent way. Our scheme applies also to infinite-dimensional quantum systems, and does not require  ideal projective measurements. A related scheme \cite{huang2023quantum} was recently proposed in a different context, focusing on    the study of non-Hermitian quantum mechanics and without  making a connection to weak values.

To estimate the weak value  $W (A|\rho_{\rm in}, \rho_{\rm fin})$ in Eq. (\ref{mixedWV}) we employ  two copies of the system, initialized in the states $\rho_{\rm in}$ and $\rho_{\rm fin}$, respectively. To measure the observable $A$, we allow for a generally noisy measurement device, described by a   positive operator-valued measure  (POVM), that is, a collection of operators $(P_j)_{j=1}^N$ satisfying the conditions  $P_j\ge  0 \, \forall j$ and $\sum_j  P_j  = I$.  The outcomes of the measurement are associated to the possible values $(x_j)_{j=1}^N$ of the observable $A$. We take the measurement to be unbiased, meaning  that the expectation value of $A$ on an arbitrary quantum state $\rho$ is equal to the average of the values     $(x_j)_{j=1}^N$  with respect to the probability distribution $p(j|\rho) :  =  \Tr  [  P_j  \rho]$, $j\in  \{1,\dots,  N\}$.   In formula,
\begin{align}
\Tr[A  \rho]  =  \sum_j  \,  x_j\,  \,  \Tr[P_j \rho]  \qquad \forall \rho  \,  ,
\end{align} 
or, more compactly, 
\begin{align}\label{expansion}
A  =  \sum_j  \,  x_j\,  P_j   \, .
\end{align} 
An example of POVM that satisfies  the unbiasedness condition (\ref{expansion}) is a  noisy measurement of $A$. Given the canonical spectral decomposition $A  =  \sum_{j=1}^N   a_j\,  Q_j$ where $(a_j)_{j=1}^N$ are $N$ distinct eigenvalues, and $(Q_j)_{j=1}^N$ are the projectors on the corresponding eigenspaces, a noisy measurement of the observable $A$ can be defined as a POVM with operators $P_j= (1-p)  \,  Q_j  +   p\,  \lambda_j \,  I$,  where $p\in  [0,1)$ is a probability quantifying the amount of noise, $(\lambda_j)_{j=1}^N$ is a probability distribution, and $j\in  \{1,\dots, N\}$.     In this case, Eq. (\ref{expansion}) is satisfied by the assignment    $x_j   :=  (a_j   -  p \,  \overline a)/(1-p)$, with $\overline a: =  \sum_{j=1}^N   \,  \lambda_j  \, a_j$.

The operators $P_j$ can be used to define a function from measurement outcomes to complex numbers, given by  
\begin{align}\label{WVdist}
q (j|  \rho_{\rm in},  \rho_{\rm fin})   :=   \Tr[ P_j \, \rho_{\rm in}  \, \rho_{\rm fin}]   \,.
\end{align}
Mathematically, this function is a complex measure \cite{Axler2020Measure},    normalized as $\sum_{j} q (j|  \rho_{\rm in}, \rho_{\rm fin}  )    =\Tr  [  \rho_{\rm in} \, \rho_{\rm fin}]$.  We call this measure the {\em weak value  (WV) measure}.  

When condition Eq. (\ref{expansion}) is satisfied,  the weak value (\ref{mixedWV}) can be written as the ratio of two expectation values with respect to the WV measure: the expectation value $\mathbb{E}_q  (X):= \sum_j\,  x_j\,  q (j|  \rho_{\rm in}, \rho_{\rm fin})$ of the random variable $X$ and the expectation value $\mathbb{E}_q  (Y):= \sum_j\,   q (j|  \rho_{\rm in}, \rho_{\rm fin})$ of the constant random variable  $Y$ with values $y_j  =  1 \, \forall j$; in formula, 
\begin{align}\label{expq}
  W  ( A \, |\, \rho_{\rm in} , \rho_{\rm fin}  )  =\frac{\mathbb{E}_q  (X)}{\mathbb{E}_q  (Y)} \, .
\end{align}

In general,  the WV measure $q(j | \rho_{\rm in}, \rho_{\rm fin})$ is not a probability measure, and therefore $\mathbb{E}_q(X)$ and $\mathbb{E}_q(Y)$  are not  proper expectation values.  Nevertheless, we now show that every  expectation value with respect to the WV measure  can be evaluated by computing  expectations  of  suitable random variables with respect to a proper probability distribution.   To this purpose, we use the following protocol.

\begin{protocol}{Sampling from the WV measure}\label{prot:1}

\textit{Inputs:} Two copies $S_1$, $S_2$ of the system,  prepared in the states $\rho_{\rm in}$ and $\rho_{\rm fin}$, respectively, an auxiliary qubit $C$ prepared in the state $|+\>\<+|$ with $|+\>=  (|0\>  +  |1\>)/\sqrt 2$, and a measurement device described by the POVM $(  P_j)_{j=1}^N$.  

\par\addvspace{\topsep}

\textit{Protocol:}
\begin{enumerate}
\item Apply the controlled {\tt SWAP} gate 
\begin{equation}
 U  = I_{S_1}\otimes I_{S_2}  \otimes |0\>\<0|_C  +  {\tt SWAP}_{S_1 S_2} \otimes |1\>\<1|_C \, , 
\end{equation}
where ${\tt SWAP}_{S_1 S_2}$ is the {\tt SWAP} gate, uniquely defined by the relation  ${\tt SWAP}_{S_1S_2}  (  |\alpha\>_{S_1} \otimes |\beta\>_{S_2} )  =  |\beta\>_{S_1} \otimes |\alpha\>_{S_2}$, for arbitrary vectors $|\alpha\>$ and $|\beta\>$.
\item Measure $S_1$ with the POVM $(P_j)_j$.
\item Measure $C$ with the four-outcome POVM $(R_c )_{c=0}^3$ with 
\begin{align}
 \nonumber R_0 = \frac 12  |+\>\<+| \,, \\
 \nonumber   R_1 = \frac 12  |-\>\<-| \, \\
 \nonumber  R_2 = \frac 12  |+i\>\<+i|\,, \\
 R_3 = \frac 12  |- i\>\<- i|\,,
\end{align}
\end{enumerate}
\end{protocol}

\noindent where $|\pm\>=  (|0\>  \pm  |1\>)/\sqrt 2$, and $|\pm i \>  =  (|0\>  \pm i|1\>)/\sqrt 2$. Physically, the POVM $(R_c)_c$ can be realized  as a random measurement, by  measuring either on the basis $\{|+\>  ,  |-\>\}$ or on the basis $\{|+i\>,  |-i\>\}$.

\vspace*{0.25cm}

\noindent \textit{Output:} The protocol produces a pair of outcomes $(j,c)$, distributed with probability 
\begin{eqnarray}
\nonumber    &&p(j,c|\rho_{\rm in},  \rho_{\rm fin})   \\ 
&&  \qquad  :  =  \Tr\left[  (P_j\otimes I \otimes R_c) \, U (\rho_{\rm in} \otimes \rho_{\rm fin} \otimes |+\>\<+|)  U^\dag  \right] \\
\nonumber     &&  \qquad  =   \frac1{8}  \Big\{  \Tr[  P_j\,  \rho_{\rm in}]  +  \Tr[  P_j\,  \rho_{\rm fin}] \\
\nonumber       &&\qquad  \quad  + \; 2\theta  (1-c)  \,  (-1)^c  \,{\rm Re}  \left(\Tr[P_j \rho_{\rm in}  \rho_{\rm fin}] \right)\\
&& \qquad  \quad - \; 2\theta (c-2) \, (-1)^c \,  {\rm Im}  \left(\Tr[P_j \rho_{\rm in}  \rho_{\rm fin}] \right)  \Big\} \, , \label{p}
\end{eqnarray}
where $\theta (t)$ is the Heaviside step function, defined as $\theta (t)  =1$ for $t \ge  0$ and $\theta (t)  =0 $ for $t<0$. The  derivation of Eq. (\ref{p}) is provided in Appendix \ref{app:protocol}.

\par\addvspace{\topsep}

\hrule

\par\addvspace{\topsep}

Expectation values with respect to the weak value measure (\ref{WVdist}) can  then be obtained from expectation values with respect to the probability distribution (\ref{p}) generated by the above protocol.  The explicit recipe is provided by the following lemma: 
\begin{lem}\label{lem:ExpValWV}
For every random variable $Z:  j\mapsto  z_j$, the expectation value of $Z$ with respect to the WV measure  $q(j|\rho_{\rm in},  \rho_{\rm fin})$ in Eq.  (\ref{WVdist}) coincides with  the expectation value of the random variable  $\widetilde Z  :  (j,c)  \mapsto  \widetilde z_{j,c}$ defined by  
\begin{align}
\widetilde z_{j,c}   :  =  2  z_j \,  (-1)^c  \, \Big[   \, \theta(1-c) - i \theta (c-2)\Big] \label{eq:ranvar}
\end{align}
with respect to the probability distribution $p(j,c|\rho_{\rm in},  \rho_{\rm fin})$ in Eq. (\ref{p}). 
\end{lem}
The proof of the lemma is provided in  Appendix \ref{app:lemma1}. This lemma, combined  with Eq. (\ref{expq}), implies  that  the weak value $W (A| \rho_{\rm in}, \rho_{\rm fin})$ can be estimated by computing the empirical average of the  random variables   $X$ and $Y$  with respect to the experimental frequencies generated by Protocol \ref{prot:1}. The sample complexity of the  protocol is provided by the following: 
\begin{theorem}\label{theo:sample_complexity}
 The weak value $W(A\, |\, \rho_{\rm in}, \rho_{\rm fin})$ of an observable $A = \sum_j x_j P_j$ with a POVM $(P_j)_j$ and real coefficients $\{x_j\}_j$ can be estimated up to a small additive error $\epsilon$ with a probability no less than $1-\delta$ by using  $K$ copies of the states $\rho_{\rm in}$ and $ \rho_{\rm fin}$, with   $K$  given by
    \begin{eqnarray}
        K &=& \frac{8\ln\Bigl(\frac{6}{\delta}\Bigr)}{\epsilon^2} \Biggl( \frac{x_{\rm max} + |W(A\, |\, \rho_{\rm in} ,  \rho_{\rm fin})|}{\Tr[\rho_{\rm in}\, \rho_{\rm fin}]} \Biggr)^2 \nonumber\\
        && + \; O\left(\frac{\ln\Bigl(\frac{1}{\delta}\Bigr)}{\epsilon} \frac{x_{\rm max} + |W(A\, |\, \rho_{\rm in}, \rho_{\rm fin})|}{(\Tr[\rho_{\rm in}\rho_{\rm fin}])^2}\right)
    \end{eqnarray}
    where $x_{\rm max} \coloneqq \max_j |x_j|$.     
    \begin{proof}
        See Appendix \ref{app:complexity}.
    \end{proof}
\end{theorem}
Note that, when the POVM   $(P_j)_{j=1}^N$ is the spectral decomposition of the observable $A$,  $x_{\max}$ is just the eigenvalue of $A$ with maximum modulus, equal to the norm $\|  A\|   =  \sup_{\||\psi\>\|=1}  \,  \<\psi  |A  |\psi\> $.

Theorem \ref{theo:sample_complexity} provides the asymptotic scaling of the sample complexity for small $\epsilon$: it identifies  the leading order in  $1/\epsilon$ for every fixed value of $\delta$ and of the overlap $\Tr[\rho_{\rm in}\rho_{\rm fin}]$.  With respect to the overlap, the scaling $1/(\Tr[\rho_{\rm in}\rho_{\rm fin}])^2$ is the same as the scaling in the protocol of Ref.~\cite{Wagner2023}. An important difference, however, is that the sample complexity in Theorem \ref{theo:sample_complexity} does not depend  on the dimension of the system.  As a consequence,  Protocol \ref{prot:1} provides a reduction of the sample complexity with respect to  the protocol proposed of Ref.~\cite{Wagner2023} whenever the observable $A$ has rank larger than one. 

From the experimental point of view, Protocol \ref{prot:1} provides a simplification with respect to the protocol of Ref.~\cite{Wagner2023}, in that  it only requires a controlled-{\tt SWAP} operation, instead of  controlled shift operations, whose basic implementation consists of a sequence of controlled {\tt SWAP}s~\cite{Wagner2023}. Since the $\tt SWAP$ gate can be implemented    as a sequence of three $\tt CNOT$ gates  \cite{lee2006cost},  the controlled $\tt SWAP$ gate can be similarly realized as a sequence of three controlled-{\tt CNOT} gates, also known as Toffoli gates. In the circuit model of quantum computing, the controlled {\tt SWAP} gate is also known as the Fredkin gate and recently has been experimentally implemented with photonic qubits \cite{patel2016quantum,wang2021experimental}.

  It is also interesting to compare the sample complexity of our protocol with the sample complexity of the traditional  measurement scheme involving weak interactions and postselection.   When the initial and final states are  pure, the sample complexity was evaluated in Refs. \cite{kofman2012nonperturbative, Wagner2023}, and was shown to scale inverse linearly with the overlap $\Tr[\rho_{\rm in}\rho_{\rm fin}]   = |\langle \psi_{\rm in}|\psi_{\rm fin}\rangle |^2$.  In this case, the weak measurement scheme appears to have lower sample complexity whenever $\rho_{\rm in}\not =  \rho_{\rm fin}$. In the mixed state case, however, the   situation is generally different.   A weak measurement scheme for measuring mixed-state weak values  was provided by Vaidman \textit{et al} \cite{Vaidman2017},  who showed that every mixed-state weak value   can be reduced to a pure-state weak value by introducing three auxiliary systems, of the same dimension of the original system $S$ and hereafter denoted by $A_1$, $A_2$, and $A_3$. Specifically, the purification recipe by Vaidman \textit{et al} can be summarized by the following formula:
  \begin{align}\label{vaidman}
  W  ( A|   \rho_{\rm in},  \rho_{\rm fin})   =  W  (A \otimes I_{A_1} \otimes I_{A_2} \otimes I_{A_3}  |   \Psi_{\rm in} ,   \Psi_{\rm fin}),  
  \end{align}
  where $I_{A_1},I_{A_2},I_{A_3}$ denotes the identity operator on the auxiliary systems $A_1,A_2,A_3$, respectively, and $\Psi_{\rm in}$ and $\Psi_{\rm fin}$ are the pure states associated to the vectors 
  \begin{align}
\nonumber   |\Psi_{\rm in}  \rangle_{S A_1A_2A_3}      & =|\Psi\rangle_{S A_1} \otimes |\Xi  \rangle_{A_2A_3}  \\  
|\Psi_{\rm fin}  \rangle_{S A_1A_2A_3}      & =|\Phi\rangle_{S A_2} \otimes |\Gamma  \rangle_{A_1A_3}\, ,
  \end{align}
where, in turn,  the vectors 
  \begin{align}
\nonumber  |\Psi\rangle_{SA_1}  &:=  \sum_{m}   \sqrt{r_m}  \,  |\psi_m\rangle_S \otimes  |m\rangle_{A_1}     \\
 |\Phi\rangle_{SA_2}  &:=  \sum_{n}   \sqrt{r_n'}  \,  |\psi'_n\rangle_S \otimes  |n\rangle_{A_2}  \, ,     
  \end{align}
  are purifications of the states $\rho_{\rm in}   =  \sum_{m}  \,  r_m\,  |\psi_m\rangle \langle \psi_m|$ and $\rho_{\rm fin}   =  \sum_{n}  \,  r_n'\,  |\psi_n'\rangle \langle \psi_n'|$,  respectively,   and  the vectors  $|\Xi\rangle_{A_2A_3}$ and $ |\Gamma\rangle_{A_1A_3}$ are defined as 
  \begin{align}
  |\Xi\rangle_{A_2A_3} &:  =   \frac 1{\sqrt d} \,  \sum_{j=1}^d   \,  |j\rangle_{A_2} \otimes |j  \rangle_{A_3} \\
   |\Gamma\rangle_{A_1A_3} &:  =    \frac 1 {\sqrt{
   \Tr[  \rho_{\rm in }  \rho_{\rm fin}]}}\sum_{m,n}    \,  \sqrt{r_m  \,  r_n'} \, \langle    \psi_n'|\psi_m\rangle \,   |m\rangle_{A_1} \otimes |n  \rangle_{A_3}\,.
  \end{align}
  The validity of Eq.  (\ref{vaidman}) can be checked straightforwardly from  the definitions.  
  With the above choice of purification,  the overlap between the initial and final states is $\Tr [  \Psi_{\rm in}  \Psi_{\rm fin}]    =  \Tr [\rho_{\rm in}  \rho_{\rm fin}]/d$.  Hence, the calculation of Refs. \cite{kofman2012nonperturbative, Wagner2023} implies that the sample complexity scales as $d/\Tr [\rho_{\rm in}  \rho_{\rm fin}]$.  Comparing this scaling with the $1/(\Tr [\rho_{\rm in}  \rho_{\rm fin}])^2$ scaling of our protocol, we obtain that  the weak measurement approach by Vaidman {\em et al} has lower sample complexity when $\Tr [\rho_{\rm in}  \rho_{\rm fin}]$ is small compared to  $1/d$, but higher sample complexity when  $\Tr [\rho_{\rm in}  \rho_{\rm fin}]$ is large compared to $1/d$.  Besides the different scalings of the sample complexity, it is also important to observe that our scheme and the scheme by Vaidman {\em et al} use different ingredients: our scheme uses access to the states $\rho_{\rm in}$ and $\rho_{\rm fin}$, while the scheme by Vaidman {\em et al} uses access to the pure state $|\Psi_{\rm in}\rangle$ and to a  measurement that projects onto the pure state $|\Psi_{\rm fin}\rangle$.   Both state and measurement include three auxiliary systems in addition to the system on which the weak value is defined, and preparing the state  $|\Psi_{\rm in}\rangle$  requires access to a purification of the state $\rho_{\rm in}$, which a strictly stronger resource compared to access to a device that prepares   $\rho_{\rm in}$.

Finally,  another interesting aspect of Protocol \ref{prot:1} is that its operational setup mirrors the mathematical expression of the weak value:  the two states $\rho_{\rm in}$ and $\rho_{\rm fin}$ in the mathematical expression $W  (A  |  \,  \rho_{\rm in},  \rho_{\rm fin})  =  \Tr[  \rho_{\rm fin} \,  A\, \rho_{\rm in}]/\Tr[\rho_{\rm fin} \rho_{\rm in}]$  correspond to  two state preparations in the setup, and the observable $A$ corresponds to the measurement of an observable in the setup.   In this aspect, Protocol \ref{prot:1} differs from other protocols, where the  observable $A$ sometimes corresponds to the generator of a unitary dynamics (as in the protocols based on weak interactions) or to the preparation of a quantum state (as in the cyclic-shift protocol of ~\cite{Wagner2023}).  Later in the paper, we will see that  the correspondence between states and observables in the mathematical expression of weak value on one hand, and state preparations and measurement devices in the laboratory on the other hand  enables a connection between bipartite quantum states and two-time states.

\section{Variants of the controlled-{\tt SWAP} protocol}  \label{sec:DoubleWV}
The controlled-{\tt SWAP} protocol  presented in the previous section lends itself to several generalizations, discussed in this section.  These generalizations allow  one to estimate another type of quantities, which we call ``double weak values.''  

\subsection{Two local measurements}
A first generalization  of Protocol \ref{prot:1} is  to measure both systems $S_1$ and $S_2$,  as illustrated in Fig. \ref{fig:second}.    Measuring two observables $A$ and $B$ on systems $S_1$ and $S_2$,   respectively, gives rise to the probability distribution  
\begin{align}
\nonumber &p(j,k,c|\rho_{\rm in}, \rho_{\rm fin}) \\
  & \quad :=\Tr\left[  (P_j\otimes Q_k \otimes R_c) \, U (\rho_{\rm in} \otimes \rho_{\rm fin}  \otimes |+\>\<+|)  U^\dag  \right] \,,  
\label{doubleprob} 
\end{align}
where  $ (P_j)_j$ and $(Q_k)_k$ are the POVMs associated to observables $A$ and $B$, respectively.

By sampling over this probability distribution, one can estimate the averages  of arbitrary random variables with respect to the complex measure  
\begin{align}
q(j,k\, |\,  \rho_{\rm in}, \rho_{\rm fin})  :  = \Tr[  P_j  \rho_{\rm in} \,  Q_k \,  \rho_{\rm fin} ] \, ,     
\end{align}
thereby estimating all  quantities of the form  
\begin{align}
\Tr  [  A \,  \rho_{\rm in} \, B \, \rho_{\rm fin}] \,,
\end{align} 
for arbitrary observables $A$ and $B$ in the linear span of $\{  P_i\}_i$ and $\{Q_j\}_j$, respectively.   

These quantities give rise to  a generalization of the notion of weak value: 
\begin{definition} 
The {\em double weak value} of a pair of observables $A \in \Lin  (\spc H)$ and $B\in \Lin  (\spc H)$ with respect to the initial state $\rho_{\rm in}$ and the final state $\rho_{\rm fin}$ is the quantity  
\begin{align}\label{doubleWV}
W_2(  A,B  \,|\, \rho_{\rm in}, \rho_{\rm fin}) : =    \frac{ \Tr  [  A\,   \rho_{\rm in}   \,  B \,  \rho_{\rm fin}]}{\Tr [  \rho_{\rm in} \, \rho_{\rm fin}]} \, .
\end{align}
\end{definition}
The double weak value generalizes the standard weak value (\ref{mixedWV}), whose expression can be retrieved from Eq. (\ref{doubleWV}) by setting $B=  I$; in short,  one has
 \begin{align}
 W(  A  \,|\, \rho_{\rm in}, \rho_{\rm fin})  =  W_2(  A,  I  \,|\, \rho_{\rm in}, \rho_{\rm fin})   \qquad \forall A ,\rho_{\rm in}, \rho_{\rm fin}  \in \Lin  (\spc H)\, .
 \end{align}
 
  The double weak value has an interesting physical interpretation.  For pure states $\rho_{\rm in}  =  |\psi_{\rm in}\>  \<\psi_{\rm in}|:=  \psi_{\rm in}$  and $\rho_{\rm fin}  =  |\psi_{\rm fin}\>\<\psi_{\rm fin}|:=  \psi_{\rm fin}$,  one has 
 \begin{align}
 W_2(  A,B  \,|\,  \psi_{\rm in}, \psi_{\rm fin})    =   W(  A  |\psi_{\rm in}, \psi_{\rm fin})  ~ W(  B  |\psi_{\rm fin}, \psi_{\rm in}) \, ,  
 \end{align}
meaning that the double weak value  is the product of the weak value of $A$ in the forward time direction, with initial state $\psi_{\rm in}$ and final state $\psi_{\rm fin}$, and the weak value of $B$ in the backward time direction, with   initial state $\psi_{\rm fin}$ and final state $\psi_{\rm in}$.

For general   mixed states $\rho_{\rm in}  =  \sum_m  \, r_m  \,  |\psi_m\>\<\psi_m|$ and $\rho_{\rm fin}  =  \sum_{n}  \, r'_{n}  \,  |\psi'_n\>\<\psi'_n|$, the  double weak value  quantifies the correlations between the weak values of $A$ in the forward time direction and the weak values of $B$ in the backward time direction; in formula: 
\begin{align}\label{W2correlations} 
W_2(  A,B  \,|\, \rho_{\rm in}, \rho_{\rm fin})  = \sum_{m,n}  \,  p (m,n)\,  W(  A  |\psi_m, \psi_n')  ~ W(  B  |\psi_n', \psi_m)  \,, 
\end{align}
where $p(m,n)$ is  the probability distribution 
\begin{align}
   p(m,n)  :  =  \frac{  r_m  r'_{n}   \,|\<\psi_m|\psi_n'\>|^2  }{  \sum_{i,j} r_i  r'_{j}   \,|\<\psi_i|\psi_j'\>|^2} \, .
\end{align}
The above expression provides an alternative way to estimate the double weak value, by sampling over pairs consisting of a pure initial state $\psi_m$ and a pure final state $\psi_n'$, according to the probability distribution $p(m,n)$.

Recalling that the weak values were originally defined in terms of weak measurement processes,  the double weak value in Eq. (\ref{W2correlations}) can be interpreted as the correlation between the quantities observed in two weak measurement processes where the roles of the pre- and post-selections are exchanged.   These correlations would appear in an exotic scenario where two different agents operate in two opposite time directions \cite{chiribella2022quantum}, with one agent preparing inputs in the past and selecting outputs in the future, and the other agent preparing inputs in the future and selecting outputs in the past. In this setting, the double weak value is the correlation between the values observed by the two agents.    It is quite remarkable that such exotic correlations can be experimentally observed through the controlled-{\tt SWAP} protocol.

Yet another way to measure the double weak value for mixed states is to use a purification approach, in a similar vein as it was proposed by  Vaidman {\em et al} for the  mixed-state weak values \cite{Vaidman2017}. Here, we show that the double weak value for mixed states can be reduced to the (standard) weak value for pure states.  To this purpose, we consider two copies of the system of interest, denoted by $S_1$ and $S_2$, respectively, and two auxiliary systems, $A_1$ and $A_2$, of  the same dimension of the system.   Explicitly, the reduction is provided by  the following relation:  
\begin{align}\label{purifieddouble}
W_2(  A,B  \,|\, \rho_{\rm in}, \rho_{\rm fin})  =   W(  A \otimes B \otimes I_{A_1} \otimes I_{A_2} \,|\,  \Gamma_{\rm in}  ,  \Gamma_{\rm fin} )   \, ,
\end{align} 
where  $\Gamma_{\rm in}$  and $\Gamma_{\rm fin}$ are  projectors onto the unit vectors 
\begin{align}
\nonumber |\Gamma_{\rm in}  \rangle  &= |\Psi\rangle_{S_1 A_1}  \otimes   |\Phi\rangle_{S_2 A_2}\\
|\Gamma_{\rm fin}  \rangle  &= |\Phi\rangle_{S_1 A_2}  \otimes   |\Psi\rangle_{S_2 A_1} \, ,
\end{align}
 $|\Psi\rangle   :=\sum_m  \sqrt{r_m}  \,  |\psi_m\rangle \otimes |m\rangle$
  and $|\Phi\rangle   :=\sum_n  \sqrt{r_n'}  \,  |\psi_n'\rangle \otimes |n\rangle$  being purifications of the states $\rho_{\rm in}   =  \sum_{m}  \,  r_m\,  |\psi_m\rangle \langle \psi_m|$ and $\rho_{\rm fin}   =  \sum_{n}  \,  r_n'\,  |\psi_n'\rangle \langle \psi_n'|$, respectively.  The validity of Eq.  (\ref{purifieddouble}) can be checked straightforwardly from the above definitions.    
  
  It is worth noting that setting  $B=I$ in Eq. (\ref{purifieddouble}) provides a new way to  reduce the mixed-state weak value  $W  (A|  \rho_{\rm in},  \rho_{\rm fin})$ to a pure-state weak value. This purification scheme is different from the  scheme by Vaidman {\em et al}, as it involves different pre- and post-selections.   Our scheme also implies a way to measure  the mixed-state weak value  $W  (A|  \rho_{\rm in},  \rho_{\rm fin})$ through weak interactions, using pre- and post-selection on the states $\Gamma_{\rm in}$ and $\Gamma_{\rm fin}$, respectively.  Interestingly, the overlap between these states is  $\Tr  [ \Gamma_{\rm in} \Gamma_{\rm fin} ]  = ( \Tr[  \rho_{\rm in}  \rho_{\rm fin} ])^2$, and therefore the sample complexity of the above weak measurement scheme has the same scaling with respect to the overlap as the sample complexity of our controlled-{\tt SWAP} protocol.  
  
 In summary, there are at least three different types of protocols to measure  the double-weak value for mixed states: (1)  our controlled-$\tt SWAP$ protocol, (2) protocols  that measure the standard weak values for the observables $A$ and $B$ and sample over different initial and final states,  and (3) protocols that measure the weak value for the observable $A\otimes B$  for pure entangled states on an extended system.

It is also interesting to observe that, when the two observables $A$ and $B$ coincide  ($A=  B$) and  the final state is pure  ($\rho_{\rm fin}  =  |\psi_{\rm fin}\>\<\psi_{\rm fin}|$), the double weak value coincides with the associated weak value defined in   Ref. \cite{kofman2012nonperturbative}  (cf. Eq. (14.19) therein). The associated weak value was shown to arise  in a non-linear theory of weak measurements \cite{kofman2012nonperturbative}, and the variant of  Protocol \ref{prot:1} discussed in this section provides an alternative way to measure them in the laboratory, by sampling from the probability distribution   (\ref{doubleprob}), in which both POVMs $ (P_j)_j$ and $ (Q_k)_k$  correspond to two (possibly different) noisy measurements of the observable $A$.

\subsection{Local measurements and general bipartite states}  

A further generalization of Protocol \ref{prot:1}   is to replace the two uncorrelated states $\rho_{\rm in}$ and $\rho_{\rm fin}$ with a single bipartite state $\rho_{\rm in, fin}$, as illustrated in Fig. \ref{fig:third}.  
 If systems $S_1$ and $S_2$ are measured with POVMs $(P_j)_j$ and $(Q_k)_k$, respectively, the  controlled-{\tt SWAP} protocol  produces a triple of outcomes $(j,k,c)$  distributed with probability
\begin{align}
p(j,k,c|\rho_{\rm in ,fin})   :=\Tr\left[  (P_j\otimes Q_k \otimes R_c) \, U (\rho_{\rm in ,fin}   \otimes |+\>\<+|)  U^\dag  \right] \,.
\end{align}

By sampling over this probability distribution, one can simulate the averages  of arbitrary random variables with respect to the complex measure  
\begin{align}
q(j,k\, |\,  \rho_{\rm in , fin})  :  = \Tr[  (P_j  \otimes  Q_k)\,  \rho_{\rm in, fin}   \, \tt SWAP   ] \, ,     
\end{align}
thereby estimating quantities of the form  
\begin{align}\label{tracedouble}
\Tr [  (A \otimes B)  \, \rho_{\rm in, fin}   \, \tt SWAP ]  
\end{align}
for arbitrary observables $A$ and $B$ for which the POVMs provide unbiased measurements $(P_j)_j$ and $(Q_k)_k$. 

As a biproduct, one can also estimate the ratios  
\begin{align}\label{pseudodouble}
 \frac{ \Tr  [  (A\otimes B)\,    \rho_{\rm in , fin}   \, \tt SWAP]}{\Tr [  \rho_{\rm in , fin}   \, \tt SWAP   ]} \, .  
\end{align}
Note that we do not use the term ``(double) weak values'' for the ratios in Eq.  (\ref{pseudodouble}).    This omission is intentional:  as we will see in the next section,    the  proper notion of (double) weak value does not, in general, coincide with the ratios in Eq. (\ref{pseudodouble}), but rather with a variant of   Eq. (\ref{pseudodouble}) where the bipartite state $\rho_{\rm in , fin}$ is subject to a partial transposition.

\subsection{General bipartite states and joint measurements}

A third generalization of the controlled-{\tt SWAP} protocol consists in  performing a joint measurement on the two copies of the system, as illustrated in Fig. \ref{fig:fourth}.    If the two copies are  initially in the bipartite state $\rho_{\rm in, fin}$ and are measured with the joint POVM $  (\Pi_j)_j$ after the controlled-{\tt SWAP} operation,  the protocol produces  a pair of outcomes $(j,c)$ distributed with probability 
\begin{align}
p(j,c|\rho_{\rm in ,fin})   :=\Tr\left[  (\Pi_j \otimes R_c) \, U (\rho_{\rm in ,fin}   \otimes |+\>\<+|)  U^\dag  \right] \,.
\end{align}
By sampling from this probability distribution, an experimenter can estimate  the expectation value of arbitrary random variables with respect to the complex measure
\begin{align}
q (j |     \rho_{\rm in ,fin})    =   \Tr  [ \Pi_j\,  \rho_{\rm in, fin} \,  {\tt SWAP}] \, ,
\end{align}  
and therefore the value of every quantity of the form  
\begin{align}\label{tracejoint}
  \Tr  [ O\,  \rho_{\rm in, fin} \,  {\tt SWAP}] \, ,
\end{align}  
where $O \in  \Lin  (  \spc H  \otimes \spc H)$ is an arbitrary observable for which the POVM $  ( \Pi_j)_j$ provides an unbiased measurement.  Again, this setup  allows one to estimate the ratios  
\begin{align}\label{pseudojoint} 
\frac{\Tr  [  O \,    \rho_{\rm in , fin}   \, \tt SWAP]}{\Tr [  \rho_{\rm in , fin}   \, \tt SWAP   ]}.
\end{align}  

In the next  section, we will compare the experimentally  accessible quantities (\ref{tracejoint}) and (\ref{pseudojoint}) with the  weak values associated to general two-time states.

\section{Two-time states}\label{sec:TwoTimeStates}

Weak values can be interpreted as expectation values with respect to a generalized type of quantum states, known as {\em two-time states} \cite{Aharonov1988, Aharonov1991, Silva2014}. In the following, we provide a brief review of the notion of two-time state.

\subsection{Two-time vectors}
The prototype  of a two-time state is the {\em two-time vector}  introduced in the seminal work of Aharonov, Bergmann, and Lebowitz  \cite{Aharonov1964}. A two-time vector   is a linear functional $\lambda:  \Lin  (\spc H) \to \C$  defined by the relation  
\begin{align}\label{purefactorized}
\lambda  ( A) : =   \< \psi_{\rm fin}|  A  |\psi_{\rm in}\> \,,  \qquad \forall A\in \Lin (\mathcal{H}) \, , 
\end{align}  
where $|\psi_{\rm in}\>\in \spc H$ and $|\psi_{\rm fin}\>\in \spc H$ are two arbitrary vectors.    Here, the functional  $\lambda$     plays the role of an unnormalized state vector in textbook quantum mechanics.   The complex number $\lambda  (A)$  is sometimes called the {\em expectation value of the observable $A$ on the two-time vector $\lambda$}.

The notion of two-time vector was later generalized by Aharonov and Vaidman \cite{Aharonov1991}, who   considered  general linear combinations of the form  
\begin{align}\label{generalized}
\lambda (A) :  = \sum_{i  =  1}^N  \<  \psi'_i|  \, A  \,  |\psi_i\> \,,  \qquad \forall A \in  \Lin (\spc H) \, ,
 \end{align}
where $N$ is a positive integer and $\{|\psi_i\>\}_{i=1}^N\subset \spc H$ and  $\{|\psi'_i\>\}_{i=1}^N \subset \spc H$ are arbitrary sets of vectors.

Mathematically, the set of two-time vectors  (\ref{generalized}) is the set of  all  linear functionals on the observables of the system.   

\begin{pror} 
The set of all two-time vectors  (\ref{generalized}) is  the  vector space   consisting of all linear functionals from  $ \Lin  (\spc H)$ to $\C$.     
\end{pror}
 \Proof For every linear functional $\lambda:   \Lin  (\spc H) \to \C$, there exists one and only one matrix   $L\in \Lin (\spc  H)$ such that $\lambda (A)   = \Tr[   L\,   A ]$.    By the singular value decomposition  \cite{Datta1995}, the matrix $L$ can be rewritten as $L =  \sum_{i=}^d  \,  |\psi_i\>\<\psi_i'|$, where $\{|\psi_i\>\}$ and   $\{|\psi_i'\>\}$ are two sets of orthogonal states. Hence, the functional $\lambda$ has the Aharonov-Vaidman form (\ref{generalized}). This proves that every linear functional is a valid two-time vector.  The converse is trivial, since every two-time vector is, by definition, a linear functional.   \qed    
 
\medskip

\subsection{Two-time matrices}

The notion of two-time state  was later been extended from pure to mixed states  by  Silva {\em et al}  \cite{Silva2014}, who developed a framework for describing pre- and post-selected ensembles of quantum states, and by Vaidman {\em et al} \cite{Vaidman2017}, who developed the connection to weak interaction schemes.  In this subsection we now review Silva {\em et al}'s  framework, using a slightly different notation  that facilitates the connection with the notion of double weak value  introduced in this paper.  

The transition from pure to mixed two-time states is similar to the transition from  state vectors to  density matrices in textbook quantum mechanics. First, one defines the matrices corresponding to pure states: 
\begin{definition}
 The {\em two-time  density matrix} corresponding to a  two-time vector $\lambda: \Lin  (\spc H) \to \C$  is the bilinear functional  $E_\lambda  :  \Lin (\spc H ) \times \Lin( \spc H) \to \C$  defined by the relation 
\begin{align}\label{Elambda}
E_{\lambda} (A,   B)    :  =     \lambda (A)   \,    \lambda^\dag (B)  \, ,  \qquad \forall A , B\in  \Lin (\spc H)\, ,
\end{align}
where $\lambda^\dag$ is the functional defined by the relation $  \lambda^\dag (A)   : =    \overline{\lambda (  A^\dag)  } \, , \forall A \in  \Lin  (\spc H)$.   We call a matrix $E_\lambda$ of the form (\ref{Elambda}) a {\em pure two-time density matrix}.   
\end{definition} 
 Here, the term  ``matrix'' refers to the fact that the observables $A$ and $B$ in Eq. (\ref{Elambda}) can be regarded as  the row and column indices  of a matrix.  Explicitly,  $E_\lambda$   can be represented by a matrix by fixing two bases $\{A_i\}_{i=1}^{d^2}$  and $\{B_j\}_{j=1}^{d^2}$   for the space $\Lin(\spc H)$, and defining the matrix elements $[E_\lambda]_{ij}:=E_\lambda (A_i,  B_j)$.  
 
 The correspondence between the two-time state $\lambda$ and the  two-time matrix $E_{\lambda}$ is analogous to the correspondence between a unit vector $|\psi\>$ and the corresponding density matrix $|\psi\>\<\psi|$. 
 It is instructive to see this correspondence in a few  examples. First, a two-time vector  of the basic form  $\lambda (A)  =  \<\psi_{\rm fin}|  A  |\psi_{\rm in}\>$    gives rise to the two-time matrix 
\begin{align}\label{matrixforproductstate}
E_\lambda   (A,B) = \<\psi_{\rm fin}|  A|\psi_{\rm in}\>  \<\psi_{\rm in}|  B|\psi_{\rm fin}\> \,.
\end{align}
More generally, a two-time vector  of the Aharonov-Vaidman form  $\lambda (A)   =  \sum_{i=1}^N  \< \psi_i'|A |\psi_i\>$  gives rise to the two-time matrix     \begin{align}\label{pure2time}
E_{\lambda}   (A, B) =  \left(\sum_{i  = 1}^N  \,  \<\psi_i'|  A|\psi_i\> \right)\, \left(  \sum_{j=1}^N    \<\psi_j|  B|\psi_j'\>  \right) \, .
\end{align}   
Note that  every pure two-time matrix $E_\lambda$  satisfies the positivity condition 
\begin{align}\label{positivity}
E_{\lambda}  (A,  A^\dag)  \ge  0  \qquad\forall A \in \Lin  (\spc H) \, .
\end{align}

\begin{definition}\label{def:2timematrix}
A {\em (generally unnormalized) two-time density matrix}   is a  bilinear functional  $\omega   :  \Lin (\spc H) \times \Lin  (\spc H) \to \C$  of the form 
  \begin{align}\label{matrixdef}
\omega   = \sum_n\,     E_{\lambda_n} \, ,
\end{align}
where $(\lambda_n)_n$ are  arbitrary linear functionals    and $(E_{\lambda_n})_n$ are two-time matrices defined as in Eq.  (\ref{Elambda}).  The two-time density matrix $\omega$ is {\em normalized} if $\omega (I, I )=  1$,  where $I$ is the identity operator on $\spc H$.
\end{definition}

Note that every two-time density matrix $\omega$ obeys the positivity condition  
\begin{align}\label{positivityomega}
\omega(A,  A^\dag)  \ge  0  \qquad\forall A \in \Lin  (\spc H) \, ,
\end{align}
which follows from Eqs.  (\ref{positivity}) and (\ref{matrixdef}).  In the next section, we will show that this condition is both necessary and sufficient for a bilinary functional $\omega$ to  be a two-time density matrix.   

 Hereafter, we will refer to $\omega (A, B)$ as the {\em expectation value} of the pair of observables $(A,B)$ with respect to the two-time matrix $\omega$. One way to estimate the expectation value $\omega (A,B)$ from experimental data is to decompose $\omega$ as in Eq. (\ref{matrixdef}) and to measure the expectation values with respect to the pure two-time matrices $E_{\lambda_n}$, which are proportional to  products of weak values.  Later in the paper, we will see another way to estimate the expectation values based on the {\tt SWAP} test. 

Before concluding this section, we provide a concrete example of a mixed two-time density matrix. Consider  two-time vectors of the form $\lambda_{mn}  (A)  =  \sqrt{ p_m p_n'} \,  \<\psi_n'| A| \psi_m\>$, where $(p_m)_m$ and $(p_n')_n$ are two probability distributions, and $\{ |\psi_m\>\}_m$  and $\{|\psi_n'\>\}$ are two orthonormal bases.  This choice  gives rise to the two-time matrix  $\omega$ defined by
\begin{align}
\nonumber \omega (A,B) &=  \sum_{m,n}   E_{\lambda_{mn}}  (A,B) \\
\nonumber &=   \sum_{m,n}   p_m p_n'\,     \<\psi_n'| A| \psi_m\>\,  \<\psi_m| B| \psi_n'\> \\
&= \Tr [ A \, \rho_{\rm in} \, B\, \rho_{\rm fin}] \, ,
\end{align}
where $\rho_{\rm in} =  \sum_m \,  p_m\,  |\psi_m\>\<\psi_m|$ and $\rho_{\rm fin}   =  \sum_n  \,  p_n'  \,  |\psi_n'\>\<\psi_n'|$ are ordinary, single-time density matrices.   Note that $\omega (I,I)  = \Tr[ \rho_{\rm in} \, \rho_{\rm fin}]$:   if the overlap $\Tr[\rho_{\rm in} \, \rho_{\rm fin}]$ is nonzero,  then the two-time matrix $\omega$ can be normalized, thus obtaining the normalized two-time density matrix $ \omega_*  :=  \omega/\omega(I,I)$.  Note that the expectation value $ \omega_* (A,B)$ coincides with the double weak value defined in Eq. (\ref{doubleWV}).  

 In the following, the set of   two-time density matrices of a system with Hilbert space $\spc H$ will be denoted by ${\sf TwoTimeD}  (\spc H)$.  Mathematically, the set of all  unnormalized two-time states is a convex  cone, {\em i.e.} it contains all convex combinations and all positive multiples of its elements.   

The set of normalized two-time density matrices will be denoted by 
\begin{align}
{\sf TwoTimeD}_*  (\spc H):  =  \Big\{   \omega \in   {\sf TwoTimeD}  (\spc H)~|~  \omega (I  , I)  = 1\Big\}\, \, . 
\end{align}
Mathematically, the set of normalized two-time density matrices is a convex set, contained in the convex cone ${\sf TwoTimeD}  (\spc H)$.

\section{Characterization of the two-time states}\label{sec:MatrRep}

  Silva {\em et al}  \cite{Silva2014} showed that  (generally unnormalized) two-time states are in one-to-one correspondence with (generally unnormalized)  bipartite density matrices.  We now make this correspondence  explicit, providing an expression that connects two-time states  to  the controlled-{\tt SWAP} protocol introduced earlier in the paper.

Our main result is  the following characterization:    
\begin{theorem}\label{theo:characterization}
For every bilinear functional $\omega :  \Lin  (\spc H) \times \Lin  (\spc H) \to  \C$, the following statements are equivalent:
\begin{enumerate}  
\item $\omega$ is a two-time density matrix.
\item $\omega (A,  A^\dag)  \ge  0 $ for every $A \in \Lin  (\spc H)$.
\item There exists a positive operator $P_\omega  \in  \Lin  (  \spc H \otimes \spc H)$ such that \begin{align}\label{2timestate}
\omega  (A,  B) =  \Tr  \left[    P_\omega^{T_{2}}\,  {\tt SWAP} \, (A \otimes B  )  \right]    \qquad \forall A,  B  \in  \Lin  (\spc H) \,,
\end{align} 
where  $P_\omega^{T_2}$ is the partial transpose of $P_\omega$ over the second Hilbert space.   
\end{enumerate}
\end{theorem}
\Proof See Appendix \ref{app:matrixrep}. \qed

Note that the operator $P_\omega$ in Theorem \ref{theo:characterization} is uniquely defined: 
\begin{pror} For every bilinear functional $\omega:  \Lin  (\spc H) \times \Lin  (\spc H)  \to \C$, if two operators $P_\omega$ and $P_{\omega}'$ satisfy Eq.  (\ref{2timestate}), then they are  necessarily equal, namely $P_\omega=  P_\omega'$.
\end{pror} 
{\bf Proof.}   Since the product operators $A \otimes B$ are a spanning set for $\Lin  (\spc H  \otimes \spc H)$, the condition  $\Tr  \left[    P_\omega^{T_{2}}\,  {\tt SWAP} \, (A \otimes B  )  \right]     =  \Tr  \left[    P_\omega^{\prime \, T_{2}}\,  {\tt SWAP} \, (A \otimes B  )  \right]  \, \forall A , B \in\Lin  (\spc H)$ implies $P_\omega^{\prime \, T_{2}}\,  {\tt SWAP}   =  P_\omega^{T_{2}}\,  {\tt SWAP}$. Multiplying by $\tt SWAP$ and taking the partial transpose  on both sides of the equality,  we then obtain  $P_\omega'  = P_\omega$.  \qed      

Theorem \ref{theo:characterization}  establishes  a one-to-one correspondence  between the convex cone of two-time states and the convex cone of positive bipartite matrices.  We now analyze the correspondence further,  by characterizing the structure of the {\em normalized} two-time states  and showing that they are in one-to-one correspondence with a convex subset of bipartite density matrices.
\begin{theorem}\label{theo:normalized}
A bilinear functional $\omega:  \Lin  (\spc H) \times \Lin( \spc H)\to \C$    
is a normalized two-time state  if and only if  it is of the form 
\begin{align}\label{normalized/rho}
\omega  (A,  B)    =     \frac{\Tr  \left[      \rho_{{\rm in, fin}}^{T_2}  \,  {\tt SWAP}\,  (A \otimes B)   \right]}{ \Tr  \left[\rho_{{\rm in, fin}}^{T_2} \,  {\tt SWAP}  \right]   }    \, ,  \qquad \forall A,  B \in  \Lin (\spc H) \, ,
\end{align}
where $\rho_{{\rm in, fin}} \in  \Lin  (\spc H\otimes \spc H)$  is a normalized density matrix such that $\Tr  [\rho_{{\rm in, fin}}^{T_2} \,  {\tt SWAP} ]  \not  = 0$.
\end{theorem}
\Proof See Appendix \ref{app:matrixrepnorm}. \qed

Note that, if the state $\rho_{{\rm in, fin}}$  in Theorem \ref{theo:normalized} is of the product form $\rho_{{\rm in, fin}} =  \rho_{\rm in}  \otimes  \rho_{\rm fin}^T$, then  the expectation value $\omega  (A,  B)$  in Eq. (\ref{normalized/rho})  coincides with the double weak value defined in Eq. (\ref{doubleWV}), and therefore can be estimated using our controlled-{\tt SWAP} protocol.  More discussion on  the estimation of expectation value of general two-time states using (variants of) the  controlled-{\tt SWAP} protocol will be provided in the next section.

It is worth noting the relation  
\begin{align}
\Tr \left[\rho^{T_2} \,  {\tt SWAP}\right] =   \bb  I|   \,  \rho  \, |I\kk \, ,   \qquad \forall    \rho   \in  \Lin(\spc H \otimes \spc H),
\end{align}
where   $    |I\kk :  =  \sum_{i=1}^d   |i\> \otimes |i\>$ is the canonical unnormalized maximally entangled state.  Using this relation, we can see that  Theorem \ref{theo:normalized} establishes a one-to-one correspondence between the set of normalized two-time density matrices and the subset of bipartite density matrices defined  by   
\begin{align}
\nonumber {\sf D}_*  (\spc H\otimes \spc H)  :  =   & \Big\{   \rho_{{\rm in, fin}} \in  \Lin  (\spc H\otimes \spc H)  ~|~  \rho_{{\rm in, fin}}  \ge  0 \, ,\\
\nonumber  &\qquad \qquad  \qquad \qquad  \qquad   \Tr [\rho_{{\rm in, fin}}]  = 1\, ,  \\ 
 &\qquad \qquad  \qquad \qquad  \qquad  \bb I| \rho_{{\rm in, fin}}   |I\kk   >  0  \Big\} \, .
\end{align} 
Note also that the correspondence between two-time states and bipartite matrices is   a homeomorphism (that is, it is invertible and continuous), as it is given by the linear fractional transformation  $\rho \mapsto \omega_\rho$, where $\omega_\rho$ is the bilinear functional defined by 
\begin{align}\label{omegarho}
\omega_\rho (A,  B)  :=  \frac{\Tr  [  \rho^{T_2}  \,  {\tt SWAP} \,  (A\otimes B)]}{\Tr  [  \rho^{T_2}  \,  {\tt SWAP} ]}  \, .
\end{align}
Summarizing,  we have the following 
\begin{cor} 
The sets  of normalized two-time states and bipartite density matrices with nonzero overlap with the maximally entangled state are homeomorphic; in formula,    ${\sf TwoTimeD}_*  (\spc H) \simeq   {\sf D}_*  (\spc H\otimes \spc H)$. 
\end{cor}

Note that the correspondence between two-time density matrices and bipartite density matrices  mapping is non-linear and it does not preserve convex combinations.    Nevertheless,  it is linear fractional, and therefore it maps convex combinations into convex combinations, although  with generally different weights. Explicitly, the mapping $\rho  \mapsto  \omega_\rho$  in Eq. (\ref{omegarho}) maps a convex combination of density matrices, say  $\rho=  \sum_i\,  p_i\,   \rho_i$, into a convex combination of two-time states $\omega_\rho   =  \sum_i  q_i\,  \omega_{\rho_i}$, with $q_i  =    p_i    \, \<\Phi|   \rho_i\,  |\Phi\> /(\sum_j    p_j\,  \<\Phi|   \rho_j\,  |\Phi\>  )$.    This condition implies that the mapping $\rho \mapsto \omega_\rho$ maps pure bipartite states into extreme points of the set of two-time states, and {\em vice-versa}.

\medskip

\section{Two-time states and the controlled-{\tt SWAP} protocol}\label{sec:TwoTimeSWAP}

Theorem \ref{theo:normalized} reveals a fundamental connection between two-time states and the controlled-{\tt SWAP} protocol presented earlier in the paper.   As we saw in Eq. (\ref{pseudodouble}),  the controlled-{\tt SWAP} protocol allows an experimenter to estimate any quantity of the form  
\begin{align}\label{pseudoomega}
\widetilde \omega_\rho (A, B)   :=  \frac{  \Tr  [        \rho  \,  {\tt SWAP} \, (A \otimes B)  ]}{ \Tr  [       \rho  \,  {\tt SWAP} ] } \, ,
\end{align}
for every  pair of  observables $(A,B)$ and every bipartite quantum state $\rho$.  

The difference between the  quantities  (\ref{pseudoomega}) and the expectation values on two-time states (\ref{normalized/rho}) is the presence of the partial transpose on the second Hilbert space.   Physically,   the  partial transpose can be interpreted as the signature of  the difference between the spatial correlations accessible with the controlled-{\tt SWAP} protocol and the time correlations associated to two-time states.   

By comparing Eqs.  (\ref{pseudoomega}) and (\ref{normalized/rho}) we can also have a clear view of the strengths and limitations of the controlled-{\tt SWAP} protocol.  First, if the  density matrix $\rho$ is invariant under partial transpose, an experimentalist who has access to two systems in the joint state $\rho$ can directly estimate the expectation values on the two-time state $\omega_\rho$, by using the controlled-{\tt SWAP} protocol.

More generally, if the density matrix is positive under partial transpose (PPT) \cite{Peres1996, Horodecki1996}, the controlled-{\tt SWAP} protocol can provide an estimate of the expectation values on the two time-state  $\omega_\rho$,  if the experimenter is given access to the quantum state $\rho^{T_2}$.  This is the case of the protocols shown earlier in the paper, where we saw how to estimate the weak values (\ref{pureWV}) and (\ref{mixedWV}), which correspond to the expectation values of the observables    $(A,   I)$ on the two-time states   
$\omega_{\psi_{\rm in} \otimes \psi_{\rm fin}^T}$ and $\omega_{\rho_{\rm in} \otimes \rho_{\rm fin}^T}$, respectively.     Similarly, the double weak value  $\Tr[\rho_{\rm fin} \, A\,  \rho_{\rm in} \,  B]/\Tr[\rho_{\rm fin}\, \rho_{\rm in}]$ is the expectation value of the   observables $(A ,B)$ with respect to the two-time state $\omega_{\rho_{\rm in} \otimes  \rho_{\rm fin}^T}$.

  In contrast, the controlled-{\tt SWAP} protocol does not provide, in general, an estimate of the expectation values on a two-time state when the density matrix $\rho$  is not PPT.   One way to circumvent this problem would be to approximate the partial transpose operation with a physical process, such as the optimal universal transpose map 
   \cite{Horodecki2001, Buscemi2003}. 
   However,  the estimation  protocol resulting from this approach would be dimension-dependent:   indeed,  optimal universal transpose is  the completely positive trace-preserving map given by
   \begin{align}\label{transpose}
   \map T(\rho) =  \frac{\rho^{\rm T}  +   I}{d+1} \, ,
     \end{align} 
     and the $1/(d+1)$ factor implies  that the  sample complexity for the estimation of the expectation values $\omega_{\rho}(A, B)$     grows linearly with $d$.  
   
   Another approach is to  introduce an auxiliary system, and to reduce two-time states associated to non-PPT density matrices to extended two-time states associated to density matrices that are PPT with respect to the bipartition between pre- and post-selected systems. This approach can be easily illustrated   in the pure state case. In this case,  Aharonov and Vaidman  \cite{Aharonov1991}  showed that every two-time vector  $\lambda (A) :  = \sum_{i  =  1}^N  \<  \psi'_i|  \, A  \,  |\psi_i\>$  can be obtained by introducing an auxiliary system of dimension $N$, and a joint two-time vector  $\Lambda  (\cdot)  :=  \<\Psi_{\rm fin}|  \, \cdot \, |\Psi_{\rm in}\>  $ corresponding to the  initial vector $|\Psi_{\rm in}\>   =  \sum_i \,  |\psi_i\> \otimes  |i\> \in \spc H\otimes \spc H_{\rm aux}$ and  final vector $|\Psi_{\rm fin}\>   =  \sum_i\,  |\psi_i'\> \otimes  |i\> \in \spc H\otimes \spc H_{\rm aux}$, where $\spc H_{\rm aux}$ is the Hilbert space of the auxiliary system and $\{ |i\>\}_{i=1}^N$ is an orthonormal basis for $\spc H_{\rm aux}$.  With this definition, one has the relation
\begin{align}\label{extended}
\lambda (A)  =  \Lambda  (A\otimes I_{\rm aux}) \, ,  \qquad \forall A \in \Lin (\spc H) \, ,
\end{align}
meaning that the expectation value of the observable $A$ on the two-time vector $\lambda$ coincides with the expectation value of the  observable $A\otimes I_{\rm aux}$  on the the joint two-time vector $\Lambda$.     In turn, the two-time vector $\Lambda$  consists just in a pre-selection and a post-selection to pure states, and therefore it can be reproduced by a product (and therefore PPT) state.  Explicitly, we have 
\begin{align}\label{Lambdaomega}
\frac{\Lambda (A\otimes I_{\rm aux})}{\Lambda   (I \otimes I_{\rm aux}  )  }      =  \omega_{  \Psi_{\rm in}\otimes \Psi_{\rm fin}^T  }      (A \otimes I_{\rm aux}  \otimes I \otimes I_{\rm aux}) \, , 
\end{align}
where $\omega_{  \Psi_{\rm in}\otimes \Psi_{\rm fin}^T  }$, defined as in Eq. (\ref{normalized/rho}), is a normalized two-time state in ${\sf T}_*    (\spc H \otimes \spc H_{\rm aux} \otimes \spc H\otimes \spc H_{\rm aux})$.    Combining Eqs.  (\ref{extended}) and (\ref{Lambdaomega}),  we can see that the  normalized expectation value $\lambda  (A)/\lambda(I)$ can be estimated by the controlled-{\tt SWAP} protocol, if the experimenter has access to the pure quantum states  $\Psi_{\rm in}$  and $\Psi_{\rm fin}^T$.    

The above argument can be extended from two-time vectors to two-time matrices.  For example, the expectation values with respect to the two-time matrix $E_{\lambda}$ associated to the two-time vector $\lambda$ in the previous paragraph can be computed as  
\begin{align}\label{ultima}
\frac{E_\lambda  (  A,  B  )}{E_\lambda (I,  I)} =  \omega_{  \Psi_{\rm in}\otimes \Psi_{\rm fin}^T  }   (A   \otimes I_{\rm aux} ,   B    \otimes I_{\rm aux} ) \, . 
\end{align}
 Eq. (\ref{ultima})  means that the expectation value on the l.h.s. can be estimated by the controlled-{\tt SWAP} protocol, if the experimenter has access to the pure quantum states  $\Psi_{\rm in}$  and $\Psi_{\rm fin}^T$.

Overall, the extension  approach has the benefit of being dimension-independent, but  it requires either the two-time state $\omega$ to be known, or the  appropriate PPT state of the system and the auxiliary system to be provided to the experimenter.

\section{Conclusions and outlook}\label{sec:concl}

This paper provided a dimension-independent scheme for estimating weak values of arbitrary observables. The scheme  is based on  the controlled-{\tt SWAP} gate, and generates a probability distribution that can be used to sample from a complex measure underlying the weak values of interest. 
Crucially, the scheme is modular: changing the initial and final states of the system in the mathematical expression of the weak value  corresponds  to changing two preparation devices  in the experimental setup, while changing the observable in the mathematical  expression corresponds to changing a measurement device  in the experimental setup.   

The structure of the controlled-{\tt SWAP}  scheme provides several insights into theory of two-time states.  In particular, we have derived an alternative expression for  two-time states, which provides an explicit characterization of the correspondence between two-time states and (a subset of) bipartite density matrices.   Using this expression, we    showed that the controlled-{\tt SWAP} protocol can be used to estimate the expectation values of all two-time states corresponding to PPT density matrices  (including, of course, the product density matrices corresponding to the usual definition of weak values).   For two-time states corresponding to  non-PPT density matrices,  the controlled-{\tt SWAP} protocol can still be used if the experimenter is  given access to an extendend quantum states involving a pair of auxiliary systems. 

Due to the ubiquitous presence of weak values in quantum mechanics, our estimation  scheme can be useful in several applications \cite{Dressel2014}, including {\em e.g.}  estimating quantum sensitivity   \cite{Aharonov1988,wiseman2002weak},    coherence \cite{budiyono2023quantifying}, and asymmetry \cite{budiyono2023operational}.    An interesting open question for future research is whether dimension-independent sampling schemes like ours could be constructed for the estimation of Kirkwood-Dirac (KD) quasiprobability distributions, an important type of complex measures that often arise in quantum information and foundations~\cite{Dressel2015, YungerHalpern2018, Lostaglio2022, Budiyono2023, arvidssonshukur2024properties}.   KD distributions have a close  connection with weak values, and their negativity provides a signature of quantum contextuality in Spekkens' formulation~\cite{Spekkens2005}. In turn, KD distributions have found numerous applications beyond quantum foundations, including quantum metrology~\cite{ArvidssonShukur2020, Jenne2021, LupuGladstein2022}, condensed matter physics~\cite{YungerHalpern2018, GonzalezAlonso2019, GonzalezAlonso2022}, and thermodynamics~\cite{Lostaglio2022}.  The approach of  Wagner {\em et al}  \cite{Wagner2023} provides a way to estimate the value of the KD distribution at every fixed point. The open question is whether there exist ways to simulate sampling from the KD distribution, in analogy to the scheme for sampling    from the weak distribution proposed in this paper.  

Another interesting direction concerns the theory of two-time states and its relation of the study of causality in quantum theory \cite{chiribella2009beyond, Chiribella2013, Oreshkov2012, Brukner2014, Barrett2019, Barrett2021}. Recent findings have suggested that anomalous weak values of observables shared between several parties can assist in witnessing the causal relationships between the parties' laboratories  \cite{Abbott2019}. Moreover, two-time states (and multiple-time states in general) themselves can carry information about the underlying causal structure \cite{Silva2017} and can be used to witness indefinite causal order of operations \cite{Liu2023ICO}. Finally,  protocols exploiting controlled causal order of operations in a protocol known as the quantum SWITCH \cite{chiribella2009beyond, Chiribella2013}  have been shown to be useful for efficient estimation of quantities that can be connected with weak values and KD distributions, such as out-of-time-correlators \cite{Swingle2016} and incompatibility of quantum observables \cite{Gao2022}. Therefore, an interplay between weak values, KD distributions,  and indefinite causal order in quantum SWITCH appears as a promising direction for future investigations.

\begin{acknowledgements}
We  thank Nicole Yunger Halpern and Billy F. Braasch, Jr. (Quantum Steampunk Laboratory, University of Maryland) for many fruitful discussions and useful comments that helped us improve the manuscript. 
   This work was supported by funding from the Hong Kong Research Grant Council through grant no.\ 17307520,  the Senior Research Fellowship Scheme SRFS2021-7S02, and the Theme-Based Research Scheme T45-406/23-R, from the Chinese Ministry of Science and Technology through grant 2023ZD0300600,  and from  the John Templeton Foundation through grant  62312, ``The Quantum Information Structure of Spacetime'' (qiss.fr). The opinions expressed in this publication are those of the authors and do not necessarily reflect the views of the John Templeton Foundation. This research was funded in whole or in part by the Austrian Science Fund (FWF) 10.55776/PAT4559623. For open access purposes, the author has applied a CC BY public copyright license to any author-accepted manuscript version arising from this submission. Research at the Perimeter Institute is supported by the Government of Canada through the Department of Innovation, Science and Economic Development Canada and by the Province of Ontario through the Ministry of Research, Innovation and Science.
\end{acknowledgements}

\bibliography{main}

\appendix

\section{Derivation of Eq. (\ref{p})} \label{app:protocol}
In the following, we will use the short-hand notation $\rho :  = \rho_{\rm in}$ and $\rho'  :  = \rho_{\rm fin}$.

At the beginning of the protocol, the two copies of the system ($S_1$ and $S_2$)  and the control qubit are in the  product state 
\begin{equation}
    \omega = \rho \otimes \rho' \otimes |+\rangle\langle +|.
\end{equation}
Application of the controlled-{\tt SWAP} gate $U$ produces the new state
\begin{eqnarray}
    \nonumber U\omega U^\dagger &=& \frac{1}{2} \Bigl\{ \rho \otimes \rho' \otimes |0\rangle\langle 0| + \rho' \otimes \rho \otimes |1\rangle\langle 1| \\
    \nonumber && +\;  (\rho \otimes \rho'){\tt SWAP} \otimes |0\rangle\langle 1|  \\
    && +\;  {\tt SWAP} (\rho \otimes \rho') \otimes |1\rangle\langle 0| \Bigr\} \, .
\end{eqnarray}
 A measurement of the first system and auxiliary qubit with the POVM $(P_j)_j$ and $(R_c)_c$, respectively, produces a pair of outcomes $(j, c)$ distributed with probability
\begin{eqnarray}
    \nonumber p(j, c |\rho, \rho') &=& \frac{1}{2}\operatorname{Tr}\Bigl[ P_j \rho \otimes \rho' \otimes R_c |0\rangle\langle 0| \\
    \nonumber &&+ \; P_j \rho' \otimes \rho \otimes R_c |1\rangle\langle 1| \\
    \nonumber &&+ \; (P_j \otimes I)(\rho \otimes \rho'){\tt SWAP} \otimes R_c |0\rangle\langle 1| \\
    \nonumber &&+ \; (P_j \otimes I) {\tt SWAP} (\rho \otimes \rho') \otimes R_c |1\rangle\langle 0|  \Bigr] \, .
    \end{eqnarray} 
    Using the relations 
    \begin{align}
    \<0|  R_c|  0\>  =  \<1|  R_c|1\>   =  \frac 14    \qquad \forall c\in \{0,1,2,3\}
    \end{align}
    and 
     \begin{align}
    \<0|  R_c|  1\>  =  \frac {(-1)^c}4 \,  \Bigl( \theta(1-c) + i\theta(c-2)\Bigr)     \qquad \forall c\in \{0,1,2,3\}   
    \end{align}
    we then obtain  
  \begin{eqnarray}  
    \nonumber   p(j, c |\rho, \rho')    &=& \frac{1}{8}\Bigl\{ \operatorname{Tr}[ P_j \rho ] +  \operatorname{Tr}[ P_j \rho' ] \\
    \nonumber && + \; (-1)^c \Bigl( \theta(1-c) + i\theta(c-2)\Bigr)\\
    \nonumber && \cdot \; \operatorname{Tr}\Bigl[ (P_j \otimes I) (\rho \otimes \rho') {\tt SWAP} \Bigr] \\
    \nonumber && + \; (-1)^c \Bigl( \theta(1-c) - i\theta(c-2)\Bigr) \\
    && \cdot \; \operatorname{Tr}\Bigl[(P_j \otimes I){\tt SWAP}(\rho \otimes \rho') \Bigr] \Bigr\}. \label{app:eq:prob_swap}
\end{eqnarray}
Finally, using the relation $\Tr[     (A\otimes B)\,  {\tt SWAP} ]  = \Tr[AB]\, , \forall A,  B\in  \Lin (\spc H)$, we obtain 
\begin{eqnarray}
    \nonumber p(j, c |\rho, \rho') &=& \frac{1}{8}\Bigl\{ \operatorname{Tr}[ P_j \rho ] +  \operatorname{Tr}[ P_j \rho' ] \\
    \nonumber && + \; (-1)^c \theta(1-c) \Bigl(\operatorname{Tr}[P_j \rho\rho'] + \operatorname{Tr}[\rho  P_j \rho'] \Bigr)\\
    \nonumber && + \; i (-1)^c \theta(c-2)\Bigl(\operatorname{Tr}[P_j \rho\rho'] - \operatorname{Tr}[\rho P_j \rho'] \Bigr) \Bigr\} \\
    \nonumber &=& \frac{1}{8}\Bigl\{ \operatorname{Tr}[ P_j \rho ] +  \operatorname{Tr}[ P_j \rho' ] \\
    \nonumber && + \; 2(-1)^c \theta(1-c) \operatorname{Re}\operatorname{Tr}[P_j \rho\rho']\\
    && - \; 2 (-1)^c \theta(c-2)\operatorname{Im}\operatorname{Tr}[P_j \rho\rho']\Bigr\},
\end{eqnarray}
where the second equality follows from the identity $\Tr[  \rho P_j   \rho']  =  \Tr  [  \rho'\rho  P_j]   =  \Tr[   ( P_j \rho \rho'  )^\dag]  = \overline {\Tr  [   P_j \rho \rho']}$.    This concludes the proof of Eq. (\ref{p}).

\section{Proof of Lemma \ref{lem:ExpValWV}}\label{app:lemma1}  
In the following, we will use the short-hand notation $\rho :  = \rho_{\rm in}$ and $\rho'  :  = \rho_{\rm fin}$.

The proof of Lemma \ref{lem:ExpValWV} is a straightforward calculation: we only need to calculate the expectation value of the  random variable $\widetilde Z$ in Eq. (\ref{eq:ranvar}) with respect to the probability distribution $p(j,c|\rho, \rho')$.  Explicitly, the expectation value is 
\begin{eqnarray}
    \nonumber \mathbb{E}_p[\widetilde Z] &=&  \sum_{j,c}  \,  \widetilde   z_{j,c}  \, p(j,c|  \rho,  \rho') \\
   \nonumber  &=&
    2\sum_{j,c} z_j(-1)^c \Bigl[ \theta(1-c) - i\theta(c-2)\Bigr] p(j,c |\rho, \rho') \\
    \nonumber &=& \frac{1}{4} \sum_j z_j \Bigl( \operatorname{Tr}[P_j \rho] + \operatorname{Tr}[P_j \rho'] \Bigr) \\
    \nonumber && \cdot \; \sum_c (-1)^c \Bigl[ \theta(1-c) - i\theta(c-2)\Bigr] \\
    \nonumber && + \; \frac{1}{2} \sum_c (-1)^{2c} \theta(1-c) \Bigl[ \theta(1-c) - i\theta(c-2)\Bigr] \\
    \nonumber && \cdot \; \sum_j z_j \operatorname{Re}\operatorname{Tr}[P_j \rho\rho'] \\
    \nonumber && - \; \frac{1}{2} \sum_c (-1)^{2c} \theta(c-2) \Bigl[ \theta(1-c) - i\theta(c-2)\Bigr] \\
    \nonumber && \cdot \; \sum_j z_j \operatorname{Im}\operatorname{Tr}[P_j \rho\rho'] \\
    \nonumber &=& \sum_j z_j \Bigl(\operatorname{Re}\operatorname{Tr}[P_j \rho\rho'] + i\operatorname{Im}\operatorname{Tr}[P_j \rho\rho'] \Bigr) \\
    \nonumber &=& \sum_j z_j q(j|\rho,\rho') \\
    &=& \mathbb{E}_q[Z] \, .
\end{eqnarray}
In summary, the expectation value of the random variable $\widetilde Z$ with respect to the probability distribution $p(j,c|\rho, \rho')$ is equal to the expectation value of the random variable $Z$ with respect to the complex distribution $q(j|  \rho,\rho')$.  

\section{Proof of Theorem \ref{theo:sample_complexity}}\label{app:complexity}
In the following, we will use the short-hand notation $\rho :  = \rho_{\rm in}$ and $\rho'  :  = \rho_{\rm fin}$.

The proof is based on two lemmas, provided in the following. 

\begin{lem}\label{app:lemma:Nu}
    Let  $\rho$ and $\rho'$ be a pair of states, let $(P_j)_j$  be a  POVM, and let $\{x_j\}_j$ be a set of real numbers, with $x_{\max}  :  =  \max_j  |x_j |$.     Let $\nu_{\rm Re} := {\rm Re}(\Tr[\rho' A \rho])$ and $\nu_{\rm Im} := {\rm Im}(\Tr[\rho' A \rho])$ be the real and imaginary parts of the weak value of the observable $A:  =  \sum_j \,  x_j\,  P_j$, respectively.  The estimate of $\nu_{\rm Re}$ and $\nu_{\rm Im} $  obtained from  $K$ runs of Protocol \ref{prot:1} has  error at most
    \begin{equation}\label{app:eq:ErrorNu}
        \epsilon_{\nu} = 2x_{\rm max} \sqrt{\frac{1}{K}\ln\frac{2}{\delta}}
    \end{equation}
 with probability at least $1-\delta$.
    \begin{proof}
        In accordance with (\ref{expq}), $\operatorname{Tr}[\rho' A \rho] = \mathbb{E}_q[X]$, where $q$ is the WV measure defined in (\ref{WVdist}), and $X$ is a random variable taking values in $\{x_j\}_j$. Lemma \ref{lem:ExpValWV} guarantees that the probability distribution (\ref{p}) generated by Protocol \ref{prot:1} can be equivalently used to estimate $\mathbb{E}_q[X]$. In turn, it can be rewritten as
        \begin{eqnarray}
            \nonumber p(j, c|\rho,\rho') &=& \frac{1}{2}\Bigl( \theta(1-c)p_{\rm Re}(j,c|\rho,\rho') \\
            && + \; \theta(c-2) p_{\rm Im}(j,c-2|\rho,\rho')\Bigr),
        \end{eqnarray}
        with probability distributions
        \begin{eqnarray}
            \nonumber p_{\rm Re}(j,\bar{c}|\rho,\rho') &=& \frac{1}{4}\Bigl\{ \operatorname{Tr}[ P_j (\rho + \rho') ] + 2(-1)^{\bar{c}} \operatorname{Re}\operatorname{Tr}[\rho' P_j \rho]\Bigr\} \\
            \nonumber p_{\rm Im}(j,\bar{c}|\rho,\rho') &=& \frac{1}{4}\Bigl\{ \operatorname{Tr}[ P_j (\rho + \rho') ] - 2(-1)^{\bar{c}} \operatorname{Im}\operatorname{Tr}[\rho' P_j \rho]\Bigr\},
        \end{eqnarray}
        where $\bar{c} \in \{0,1\}$. 
        
        Now, recall  that  the POVM  $(R_c)_c$ in Protocol \ref{prot:1} can be obtained by  randomly choosing between the projective measurements $(2R_0, 2R_1)$ and $(2R_2, 2R_3)$,  which give rise to the probability distributions  $p_{\rm Re}(j,\bar{c}|\rho,\rho')$ and $p_{\rm Im}(j,\bar{c}|\rho,\rho')$,   respectively. Then, a straightforward calculation demonstrates that $\mathbb{E}_{p_{\rm Re}}[\widetilde X] = \nu_{\rm Re}$ and $\mathbb{E}_{p_{\rm Im}}[-\widetilde X] = \nu_{\rm Im}$, so that
        \begin{equation}\label{app:eq:ReImWV}
            \mathbb{E}_q[X] = \mathbb{E}_{p_{\rm Re}}[\widetilde X] + i\mathbb{E}_{p_{\rm Im}}[-\widetilde X].    
        \end{equation}
        In turn, $\widetilde X$ is a random variable associated with the corresponding two-outcome projective measurement of auxiliary qubit and taking values in $\{\widetilde x_{j,\bar{c}}\}_{j,\bar{c}}$, where $\widetilde x_{j,\bar{c}} = (-1)^{\bar{c}} x_j$, and $j$ and $\bar{c}$ are the measurement outcomes. Therefore, for $K$ runs of Protocol \ref{prot:1}, we can consider a projective measurement $\{2R_0, 2R_1\}$ of the auxiliary qubit performed in $K/2$ runs and the measurement $\{2R_2, 2R_3\}$ performed in other $K/2$ runs.
        
        In order to estimate $\nu_{\rm Re}$, in the $k$-th run of the protocol, we associate the measurement $\{2R_0, 2R_1\}$ with a random variable $X_k \coloneqq \widetilde{X} / (K/2) $ taking values in $\{2(-1)^{\bar{c}} x_j / K\}_{j,\bar{c}}$. Hence, the estimator $X_{\rm Re} \coloneqq \sum_{k=1}^{K/2} X_k$ is an unbiased estimator for $\nu_{\rm Re}$. According to the Hoeffding's inequality~\cite{Hoeffding1963}, the probability of estimation with an additive error at least $\epsilon_\nu$ is upper-bounded as
        \begin{equation}\label{app:eq:ProbBound}
            \Pr(|X_{\rm Re} - \nu_{\rm Re}| \geq \epsilon_\nu) \leq 2\exp\left( -\frac{2\epsilon_\nu^2}{\sum_k (\max(X_k) - \min(X_k))^2} \right).
        \end{equation}
        As each of $K/2$ estimators is bounded as $-2x_{\rm max}/K \leq X_k \leq 2x_{\rm max}/K$, where $x_{\rm max} = \operatorname{max}_j |x_j|$, we obtain 
        \begin{equation}
            \sum_k (\max(X_k) - \min(X_k))^2 = \frac{8x_{\rm max}^2}{K}.
        \end{equation}
        Therefore, the probability (\ref{app:eq:ProbBound}) is upper-bounded by
        \begin{equation}
            \Pr(|X_{\rm Re} - \nu_{\rm Re}| \geq \epsilon_\nu) \leq 2\exp\left( -\frac{\epsilon_\nu^2 K}{4x_{\max}^2} \right).
        \end{equation}
        Requiring that it does not exceed $\delta$ is equivalent to the upper bound
        \begin{equation}
            2\exp\left( -\epsilon_\nu^2 K / 4x_{\max}^2 \right) \leq \delta,
        \end{equation}
        bounding hence the estimation error as
        \begin{equation}
            \epsilon_\nu \geq 2x_{\rm max} \sqrt{\frac{1}{K}\ln\frac{2}{\delta }}.
        \end{equation}
        This means that, for $K/2$ runs of Protocol \ref{prot:1}, we can make sure that the estimation of $\nu_{\rm Re}$ is within an error $\epsilon_\nu$ with a probability no less than $1-\delta$. This relation between the error and the number of copies also holds for the estimation of the imaginary part $\nu_{\rm Im}$, since, in accordance with (\ref{app:eq:ReImWV}), it is associated with the same estimator $X_{\rm Re}$ up to an overall minus sign, which does not change the upper bound (\ref{app:eq:ProbBound}). Hence the proof.
    \end{proof}
\end{lem}

\begin{cor}\label{app:cor:Mu}
    Given a pair of states $\rho$, $\rho'$, for $K$ runs of Protocol \ref{prot:1}, the estimate of $\mu := \Tr[\rho' \rho]$ can be guaranteed to have an error at most
    \begin{equation}\label{app:eq:ErrorMu}
        \epsilon_{\mu} = 2 \sqrt{\frac{1}{K}\ln\frac{2}{\delta}},
    \end{equation}
    with probability at least $1-\delta$.
    \begin{proof}
        First, we note that $\mu = \nu_{\rm Re}$ if $A = I$. Therefore, from the $(K/2)$-round measurement outcomes for estimating $\nu_{\rm Re}$, we can simultaneously construct an unbiased estimator $Y \coloneqq \sum_{k=1}^{K/2} Y_k$ for $\mu$, where $Y_k \coloneqq 2(-1)^c / K$. Since $\mu$ is a nonnegative  number, it is not necessary to estimate its imaginary part. Applying the Hoeffding's inequality (\ref{app:eq:ProbBound}) and taking into account that $x_{\rm max} = 1$ for $\mu$, we find that the estimation error is bounded by
        \begin{equation}
            \epsilon_{\mu} \leq 2 \sqrt{\frac{1}{K}\ln\frac{2}{\delta}}.
        \end{equation}
        Therefore, for $K/2$ runs of the Protocol \ref{prot:1}, we can make sure that the estimation of $\mu$ is within a error $\epsilon_\mu$ with a probability $1-\delta$.  \end{proof}
\end{cor}

\begin{lem}\label{app:lem:ErrorWV}
    Given an observable $A = \sum_j x_j P_j$, where $(P_j)_j$ is a set of POVM elements, and a pair of states $\rho$, $\rho'$, for $K$ runs of Protocol \ref{prot:1}, the estimate of its weak value $W(A\, |\, \rho, \rho')$ can be guaranteed to have an error at most:
    \begin{equation}\label{app:eq:ErrorWV}
        \epsilon = \frac{\sqrt{2}(x_{\rm max} + |W(A\, |\, \rho, \rho')|)}{\mu/\epsilon_\mu - 1},
    \end{equation}
    where $x_{\rm max} := \operatorname{max}_j |x_j|$, $\mu = \operatorname{Tr}[\rho'\rho]$, and $\epsilon_\mu = 2\sqrt{\frac{1}{K}\ln\frac{2}{\delta}}$, with probability at least $1-3\delta$.
    \begin{proof}
        First, we we recall that $W(A\, |\, \rho, \rho') = W_{\rm Re} + i W_{\rm Im}$, where $W_{\rm Re} = \nu_{\rm Re}/\mu$ and $W_{\rm Im} = \nu_{\rm Im}/\mu$. Therefore, the corresponding estimation error is given by 
        \begin{equation}
            \epsilon = \sqrt{\epsilon_{W_{\rm Re}}^2 + \epsilon_{W_{\rm Im}}^2},
        \end{equation}
        where $\epsilon_{W_{\rm Re}}$ and $\epsilon_{W_{\rm Re}}$ are errors in estimation of the real and imaginary parts of the weak value, respectively. The error in estimation of the real part $W_{\rm Re}$ can be upper-bounded as
        \begin{align}
            \nonumber \epsilon_{W_{\rm Re}} &= \left\vert \frac{X_{\rm Re}}{Y} - \frac{\nu_{\rm Re}}{\mu} \right\vert \\
            \nonumber &= \left\vert \frac{X_{\rm Re}\mu - \nu_{\rm Re}\mu + \mu\nu_{\rm Re} - Y\nu_{\rm Re}}{Y\mu} \right\vert\\
            \nonumber &\leq \frac{|X_{\rm Re} - \nu_{\rm Re}|\mu + |\mu - Y||\nu_{\rm Re}|}{|Y|\mu}\\
            &\leq \frac{\epsilon_\nu + \epsilon_\mu|W_{\rm Re}|}{|Y|},
\end{align}
where the first inequality follows from the triangle inequality $|a + b| \leq |a| + |b|$, while the second inequality, in accordance with Lemma \ref{app:lemma:Nu} and Corollary \ref{app:cor:Mu}, follows from the upper bounds (\ref{app:eq:ErrorNu}) and (\ref{app:eq:ErrorMu}), respectively. As the latter can be given as $\mu - \epsilon_\mu \leq Y \leq \mu + \epsilon_\mu$, we can assume that the corresponding estimation error is small enough to fulfill the condition $\epsilon_\mu < \mu$, so that $|Y| \geq \mu - \epsilon_\mu \geq 0$ and, thus,
\begin{align}
    \epsilon_{W_{\rm Re}} \leq \frac{\epsilon_\nu + \epsilon_\mu|W_{\rm Re}|}{\mu - \epsilon_\mu}.
\end{align}
Comparing (\ref{app:eq:ErrorNu}) and (\ref{app:eq:ErrorMu}), it is easy to note that $\epsilon_\nu = x_{\rm max}\epsilon_\mu$. Therefore,
\begin{align}
    \epsilon_{W_{\rm Re}} \leq \frac{x_{\rm max} + |W_{\rm Re}|}{\mu/\epsilon_\mu - 1}.
\end{align}
Similarly, the error of estimating ${\rm Im}(W(A\, |\, \rho, \rho'))$ is bounded from above as
\begin{align}
    \epsilon_{W_{\rm Im}} \leq \frac{x_{\rm max} + |W_{\rm Im}|}{\mu/\epsilon_\mu - 1}.
\end{align}
Therefore, an upper bound on the total error in estimation of the weak value $W(A|\rho, \rho')$
\begin{align}
   \nonumber \epsilon 
    &\leq \frac{\sqrt{(x_{\rm max} + |W_{\rm Re}|)^2 + (x_{\rm max} + |W_{\rm Im}|)^2}}{\mu/\epsilon_\mu - 1}\\
    \nonumber &= \frac{\sqrt{2x_{\rm max}^2 + 2x_{\rm max}\left(|W_{\rm Re}| + |W_{\rm Im}|\right) + |W(A|\rho,\rho')|^2}}{\mu/\epsilon_\mu - 1}\\
    \nonumber &\leq \frac{\sqrt{2x_{\rm max}^2 + 4x_{\rm max}|W(A|\rho,\rho')| + |W(A|\rho,\rho')|^2}}{\mu/\epsilon_\mu - 1}\\
    &\leq \frac{\sqrt{2}(x_{\rm max} + |W(A|\rho,\rho')|)}{\mu/\epsilon_\mu - 1}, \label{app:eq:ErrorWVUB}
\end{align}
is valid if the bounds $|X_{\rm Re} - \nu_{\rm Re}| \leq \epsilon_\nu$, $|X_{\rm Im} - \nu_{\rm Im}| \leq \epsilon_\nu$, and $|Y - \mu| \leq \epsilon_\mu$ are satisfied. Since each bound is violated with a probability not higher than $\delta$, the violation probability of at least one of them is upper-bounded by $3\delta$. Hence, the estimation error of $W(A\, |\, \rho, \rho')$ does not exceed the upper bound (\ref{app:eq:ErrorWVUB}) with probability at least $1-3\delta$.
\end{proof}
\end{lem}

{\bf Proof of Theorem \ref{theo:sample_complexity}.} Applying Lemma \ref{app:lem:ErrorWV} and taking into account (\ref{app:eq:ErrorMu}), we can reverse (\ref{app:eq:ErrorWV}) in order to obtain an expression for the number $K$ of runs of the Protocol \ref{prot:1} in terms of the estimation error $\epsilon$,
\begin{align}
    K = \frac{4 \ln\frac{2}{\delta}}{\mu^2} \left(\frac{\sqrt{2}\left(x_{\rm max} + |W(A\, |\, \rho, \rho')|\right)}{\epsilon} + 1\right)^2.
\end{align}
Denoting $\Lambda \coloneqq \sqrt{2}\left(x_{\rm max} + |W(A\, |\, \rho, \rho')|\right) / \epsilon$ for the sake of simplicity, we obtain
\begin{eqnarray}
    \nonumber K &=& \frac{4 \ln\frac{2}{\delta} \Lambda^2}{\mu^2} \left(1 + \frac{1}{\Lambda}\right)^2\\
    \nonumber &=& \frac{4 \ln\frac{2}{\delta} \Lambda^2}{\mu^2} \left(1 + \frac{2}{\Lambda} + \frac{1}{\Lambda^2}\right)\\
    &=& \frac{4 \ln\frac{2}{\delta} \Lambda^2}{\mu^2} + O\left(\frac{\ln\frac{1}{\delta} \Lambda}{\mu^2}\right),
\end{eqnarray}
where the third equality follows from the assumption that $\Lambda \gg 1$. Finally, relabelling $3\delta$ to $\delta$ for the sake of simplicity, we conclude that
\begin{eqnarray}
    K &=& \frac{8\ln\Bigl(\frac{6}{\delta}\Bigr)}{\epsilon^2} \Biggl( \frac{x_{\rm max} + |W(A\, |\, \rho, \rho')|}{\Tr[\rho\rho']} \Biggr)^2 \nonumber\\
    && + \; O\left(\frac{\ln\frac{1}{\delta}}{\epsilon} \frac{x_{\rm max} + |W(A\, |\, \rho, \rho')|}{(\Tr[\rho\rho'])^2}\right)
\end{eqnarray}
runs of Protocol \ref{prot:1} are enough to ensure that the estimation of weak value $W(A\, |\, \rho, \rho')$ has an additive error at most $\epsilon$ with a probability at least $1-\delta$. Hence the proof. \qed

\section{Proof of Theorem \ref{theo:characterization}}
\label{app:matrixrep}

$1  \Longrightarrow 2$.  This implication was already proven by  Eq.  (\ref{positivityomega}). 

$2  \Longrightarrow 3$. A bilinear functional $\omega :  \Lin  (\spc H) \times \Lin(\spc H)$ satisfying the positivity condition $\omega (A, A^\dag) \ge 0, \, \forall A\in  \Lin  (\spc H)$ defines a scalar product on $\Lin(\spc H)$ through the relation
\begin{align}
\<  A,  B  \>_\omega  :  = \omega ( A^\dag ,  B ) \qquad \forall A,  B  \in  \Lin  (\spc H) \, .    \end{align}
With this notation, the positivity condition is equivalent to the positivity of the scalar product $\<   \cdot ~,  \cdot \>_\omega$.  

We now use the correspondence between operators in $\Lin  (\spc H)$ and bipartite vectors in $\spc H\otimes \spc H$ given by the double-ket notation $|A\kk :  =   \sum_{j=0}^{d-1}  \,  A|j\>  \otimes |j \>, \, \forall A\in \Lin  (\spc H)$.     Using this correspondence, the scalar product $\<  \cdot  ~,  \cdot \>_\omega$   can be equivalently written 
\begin{align}
\<A,  B\>_\omega  =  \bb A |  \,  P_\omega \,  |B\kk \,,  \qquad \forall A,  B \in 
\Lin (\spc H)
\end{align}
for some positive semidefinite operator $P_\omega   \in  \Lin  (\spc H \otimes \spc H)$. 

Hence, we have 
\begin{align}
\nonumber  \omega (A,  B)  &  =  \< A^\dag,   B\>_\omega   \\
\nonumber  &  =  \bb  A^\dag  |  \,  P_\omega \,  |B\kk  \\
\nonumber  &  =  \Tr  \left[    P_\omega  \,  |B\kk \bb A^\dag|  \right]\\
\nonumber  &  =\Tr  \left[  P_\omega^{T_2}     \left( |B\kk \bb A^\dag|   \right)^{T_2}  \right]\\
\nonumber  &  =\Tr  \left[  P_\omega^{T_2}    (B  \otimes  I) \left( |I\kk \bb I|   \right)^{T_2} (A  \otimes  I) \right]\\
\nonumber  &  =  \Tr   \left[  P_\omega^{T_2}  (B  \otimes  I)  {\tt SWAP}    (A  \otimes I)   \right]\\
\nonumber  &  =  \Tr   \left[  P_\omega^{T_2}  {\tt SWAP}\,   {\tt SWAP}  (B  \otimes  I)  {\tt SWAP}    (A  \otimes I)   \right]\\
&  =  \Tr   \left[  P_\omega^{T_2}  {\tt SWAP}     (A  \otimes B)   \right] \, ,
\end{align}
where $|I\kk  =   \sum_{j=0}^{d-1}  \,  |j\>  \otimes |j \>$.

$3  \Longrightarrow 1$.    Let us consider first the case where the operator $P_\omega$ is rank-one, that is $P_\omega  =  |\Gamma \>  \< \Gamma  |$ for some bipartite vector $|\Gamma\>   \in  \spc H\otimes \spc H$.    Let  us write $|\Gamma\>=  \sum_i  |\psi_i\>  \otimes |\overline \psi'_i\>$ 
for some suitable vectors $|\psi_i\>$ and $|\psi'_i\>$ in $\spc H$.  Then, we have 
\begin{align}
\nonumber  \omega  (A,B)  &   =   \Tr   \left[  P_\omega^{T_2}  {\tt SWAP}     (A  \otimes B)   \right]\\
\nonumber    &   =     \sum_{i,j} \, \Tr   \left[   \left(  |\psi_i\>\<\psi_j| \otimes |\overline \psi_i'\>\<\overline \psi_j'|\right)^{T_2}  {\tt SWAP}     (A  \otimes B)   \right]\\
\nonumber    &   =     \sum_{i,j} \, \Tr   \left[   \left(  |\psi_i\>\<\psi_j| \otimes |\psi_j'\>\<\psi_i'|\right)  {\tt SWAP}     (A  \otimes B)   \right]\\
\nonumber    &   =     \sum_{i,j} \, \Tr   \left[   \left(  |\psi_i\>\<\psi_i'| \otimes |\psi_j'\>\<\psi_j|\right)   (A  \otimes B)   \right]\\
\nonumber  &=  \sum_{i,j}  \,  \<\psi_i'|  A  |\psi_i\>  \,  \<\psi_j|  B  |\psi_j'\>  \\
&  =  \lambda (A)  \,  \lambda^\dag (B)  \, ,   \qquad \forall A,  B \in \Lin (\spc H)\, , \label{qwert}
\end{align}
where $\lambda:  \Lin  (\spc H)\to \C$ is the linear functional defined by  $\lambda( A):  = \sum_i  \<  \psi_i|  A  |\psi_i'\>$.   By definition,  $E_\lambda  (A,  B)  : = \lambda (A)  \,  \lambda^\dag (B)   =   $ is the two-time matrix associated to the two-time vector $\lambda$.   Hence,  Eq. (\ref{qwert}) implies that the functional $\omega$ is equal to the two-time matrix $E_\lambda$.  

To conclude, consider the case where the rank of  $P_\omega$ is larger than one. Since $P_\omega$ is positive semidefinite, it can be written as a sum of rank-one terms, say  $P_\omega =  \sum_n  \,  |\Gamma_n\>\<\Gamma_n|$.   By the result of the previous paragraph, we then obtain the decomposition $\omega  =  \sum_n   \,  E_{\lambda_n}$, where $\lambda_n$ is the two-time vector associated to  $|\Gamma_n\>$.  Hence, $\omega$ is a two-time matrix.  \qed

\section{Proof of Theorem \ref{theo:normalized}}
\label{app:matrixrepnorm}

Suppose that $\omega$ is a functional of the form (\ref{normalized/rho}).  Clearly, $\omega$ satisfies the normalization condition   $\omega (I, I) =1$.    Moreover, $\omega$ is of the form $\omega (A,  B) =  \Tr  [   P_\omega^{T_2}  \,  {\tt SWAP}\,  (A \otimes B)]$ with   $P_\omega: = \rho/  \Tr  [\rho^{T_2} \,  {\tt SWAP} ]$.  Note that $P_\omega$ is positive, because $\rho$ is positive and 
\begin{align}\label{swapent}    \Tr  [\rho^{T_2} \,  {\tt SWAP} ] =    \bb I |  \rho  | I \kk  \ge  0 \, ,
\end{align}
where $|I\kk  =   \sum_{j=0}^{d-1}  \,  |j\>  \otimes |j \>$. Since   $\Tr  [\rho^{T_2} \,  {\tt SWAP} ] $ is guaranteed to be non-zero, we have $\Tr  [\rho^{T_2} \,  {\tt SWAP} ]  > 0$.    Hence, $\omega$ is of the form (\ref{2timestate}), and therefore it is a (normalized) two-time state.

Conversely, suppose that $\omega$  is a normalized two-time state.  Since $\omega$ is a two-time state,  it must be of the form (\ref{2timestate}) for some positive matrix $P_\omega$.   Since $\omega$ is normalized, one has $  \Tr [  P_\omega^{T_2}  \,  {\tt SWAP}]  = \omega (I, I)  =  1$, which implies in particular $P_\omega \not  =  0$, and, since $P_\omega$ is positive, $\Tr[  P ]  >   0$.  We can then define a normalized density matrix $\rho :  =  P_\omega/\Tr[P_\omega]$ satisfying the condition $ \Tr [  \rho^{T_2}  \,  {\tt SWAP}]    =    \Tr [  P_\omega^{T_2}  \,  {\tt SWAP}]/\Tr[P_\omega]   =  1/\Tr[P_\omega]\not  =  0$.    Hence, we have 
\begin{align}
\nonumber \omega (A , B)    & =    \Tr [   P^{T_2}  \,  {\tt SWAP} \, (A \otimes B)]  \\
\nonumber &   =   \Tr  [P]\,     \Tr [    \rho^{T_2}  \,  {\tt SWAP}\,  (A \otimes B)]\\
 &=  \Tr [   \rho^{T_2}  \,  {\tt SWAP}\,  (A\otimes B)]  /  \Tr [   \rho^{T_2}  \,  {\tt SWAP}]\, .  
 \end{align}
 \qed

\end{document}